\documentclass[aps,showpacs,amssymb,amsmath,superscriptaddress,nofootinbib]{revtex4}

\usepackage{graphicx}

\usepackage{amsmath}

\usepackage{amssymb}

\def\beq{\begin{equation}}
\def\eeq{\end{equation}}

\def\bea{\arraycolsep .1em \begin{eqnarray}}
\def\eea{\end{eqnarray}}

\begin{document}

\title{Color-Superconducting Gap in the Presence of a Magnetic Field}
\author{Efrain J. Ferrer}
\author{Vivian de la Incera}
\affiliation{Department of Physics, Western Illinois University,
Macomb, IL 61455, USA}
\author{Cristina Manuel}
\affiliation{Instituto de Ciencias del Espacio (IEEC/CSIC), Campus UAB, Fac. de Ci\`encies,\\
Torre C5-Parell 2a planta, E-08193, Bellaterra (Barcelona) Spain}

\begin{abstract}

We explore the effects of an external magnetic field in the
structure and magnitude of the diquark condensate in a three
massless quark flavor color superconductor. It is shown that the
long-range component $\widetilde{B}$ of the external magnetic field
that penetrates the color-flavor locked (CFL) phase modifies its gap
structure, producing a new phase of lower symmetry. Our analysis is
performed within an NJL effective field theory with four-fermion
interactions, inspired by one-gluon exchange. Using Ritus' method,
we compute the quark propagators in the presence of a background
magnetic field, and derive the gap equations for arbitrary values of
the field. An analytical solution is found for strong magnetic
fields. A main outcome of our study is that the $\widetilde{B}$
field tends to strengthen the gaps that get contributions from pairs
of $\widetilde{Q}$-charged quarks. These gaps are enhanced through
the field-dependent density of states of the $\widetilde{Q}$-charged
quarks on the Fermi surface. Our considerations are relevant for the
study of highly magnetized compact stars.

\pacs{12.38.Aw, 12.38.-t, 24.85.+p}
\end{abstract}
\maketitle

\section{Introduction}

This article presents in full detail the study of the effects of a
strong external magnetic field in the gap structure of a color
superconductor for a system of three massless quark flavors, the
main results of which were summarized in Ref. \cite{MCFL}.

Our present knowledge of QCD at high baryonic density indicates that
quark matter might be in a color superconducting phase (for reviews,
see \cite{reviews}). This interesting possibility has originated a
flurry of research. Not only with the motivation to deepen our
knowledge about QCD, but mainly because of its astrophysical
applications. It is very likely that quark matter occupies the inner
regions of compact stars, or that even quark stars exist
\cite{Witten,qstars}. On the other hand,  it is well-known
\cite{Grasso} that strong magnetic fields, as large as $B \sim
10^{12} - 10^{14}$ G,  exist in the surface of neutron stars, while
in magnetars they are in the range $B \sim 10^{14} - 10^{15}$ G, and
perhaps as high as $10^{16}$ G \cite{magnetars}. Moreover, the
virial theorem \cite{virial} allows the field magnitude to reach
values as large as $10^{18}-10^{19}$ G. If quark stars are
self-bound rather than gravitational-bound objects, the previous
upper limit that has been obtained by comparing the magnetic and
gravitational energies could go even higher.

In this paper we investigate how a strong magnetic field can modify
color superconductivity, with the aim of further studying its
possible astrophysical implications. We will start by considering
three massless quark flavors. In this case, it is well established
that the ground state  of high-dense QCD corresponds to the CFL
(color-flavor locked) phase \cite{alf-raj-wil-99/537}.  In this
phase quarks form spin-zero Cooper pairs in the color-antitriplet,
flavor-antitriplet representation. One can now ask whether this
scenario can change when a magnetic field is switched on. Would the
external field influence the pairing phenomena? As discussed in Ref.
\cite{MCFL} and shown here in more detail, the magnetic field can
drastically affect the condensation phenomenon producing a new color
superconducting phase that we call magnetic color-flavor locking
(MCFL) phase.

To understand how a magnetic field can affect the color
superconducting pairing it is important to recall that in spin-zero
color superconductivity, although the color condensate has non-zero
electric charge, there is a linear combination of the photon and a
gluon that remains massless \cite{alf-raj-wil-99/537,
alf-berg-raj-NPB-02}. A spin-zero color superconductor may thus be
penetrated by a long-range remnant ``rotated'' magnetic field
$\widetilde{B}$. Although all the superconducting pairs are neutral
with respect to this long-range field, a subset of them is formed by
quarks of opposite rotated charges ${\widetilde Q}$. One would
expect that this set of condensates will be strengthened by the
penetrating field, since their paired quarks, having opposite
${\widetilde Q}$-charges and opposite spins, have parallel (rather
than antiparallel) magnetic moments. As will be shown in the present
paper, our intuitive expectation is confirmed by our explicit
calculations. The situation here has some resemblance to the
magnetic catalysis of chiral symmetry breaking \cite{MC,
orthonormality}, in the sense that the magnetic field strengthens
the pair formation. Despite this similarity, the way the field
influences the pairing mechanism in the two cases is quite different
as will be shown below.

Our results signalize the color superconductor as a very peculiar
kind of superconductor. Up to now this is the only physical system
where magnetism and superconductivity are not at odds with each
other, but on the contrary, the fermion-fermion pairing, which is
the hallmark of superconductivity, is reinforced by the magnetic
field.

The plan of the paper is as follows. In Sec. II we discuss the
color-flavor and Dirac structure of the gap in the presence of a
constant magnetic field. The mean-field theory in the presence of
the rotated magnetic field is introduced in Sec. III. Special
attention is devoted to obtaining the mean-field effective action in
momentum space in the presence of the external field and to the
application of the Ritus transformation \cite{Ritus:1978cj} to this
case. In Sec. IV we find the gap equations and solutions in the
presence of the $\widetilde{B}$ field. Starting from the general gap
equations for arbitrary $\widetilde{B}$ values, we analytically find
the solutions for the limiting cases of zero and strong fields
($\tilde{e}\widetilde{B} \gtrsim \mu^2$). We find that at strong
magnetic fields the gaps that get contributions from pairs of
${\widetilde Q}$-charged quarks are enhanced by the field. In Sec. V
we sketch the derivation of the low-energy effective field theory of
the Goldstone modes in the MCFL phase. The main outcomes of the
paper and discussion of the results are stated in Sec. VI.

\section{The MCFL Phase}\label{selfen-gen-struc}

As known, a degenerate quark system becomes unstable under any
quark-quark attractive interaction at the Fermi surface. This
instability leads to color-superconducting Cooper pairing. Depending
on how large the density is, there are different choices for the
interaction. For three different quark flavors, at densities higher
than the strange quark mass, the quarks can be considered
effectively massless. This system of very dense massless quarks can
form a favored state with spin-zero Cooper pairs in the
color-antitriplet, flavor-antitriplet representation (the CFL phase)
\cite{alf-raj-wil-99/537}. In compact stars where the macroscopic
volume of quark matter must be an electrically neutral color
singlet, this phase is preferred at high enough densities
\cite{alf-raj-JHEP-2002}.

An important feature of spin-zero color superconductivity is that
although the color condensate has non-zero electric charge, there is
a linear combination of the photon $A_{\mu}$ and a gluon
$G^{8}_{\mu}$ that remains massless \cite{alf-raj-wil-99/537,
alf-berg-raj-NPB-02},

\begin{equation}
\widetilde{A}_{\mu}=\cos{\theta}\,A_{\mu}-\sin{\theta}\,G^{8}_{\mu}
\ , \label{1}
\end{equation}
while the orthogonal combination
$\widetilde{G}_{\mu}^8=\sin{\theta}A_{\mu}+\cos{\theta}\,G^{8}_{\mu}$
is massive. In the CFL phase the mixing angle $\theta$ is
sufficiently small ($\sin{\theta}\sim e/g\sim1/40$). Thus, the
penetrating field in the color superconductor is mostly formed by
the photon with only a small gluon admixture.

The unbroken $U(1)$ group corresponding to the long-range rotated
photon (i.e. the $\widetilde {U}(1)_{\rm e.m.}$) is generated, in
flavor-color space, by $\widetilde {\mathbf{Q}} = \mathbf{Q} \times
\mathbf{1} - \mathbf{1} \times \mathbf{Q}$, where $\mathbf{Q}$ is
the electromagnetic charge operator. We use the conventions
$\mathbf{Q} = -\lambda_8/\sqrt{3}$, where $\lambda_8$ is the 8th
Gell-Mann matrix. Thus our flavor-space ordering is $(s,d,u)$. In
the 9-dimensional flavor-color representation that we will use in
this paper (the color indexes we are using are (1,2,3)=(b,g,r)), the
$\tilde{Q}$ charges of the different quarks, in units of $\tilde{e}
= e \cos{\theta}$, are
\begin{equation}
\label{q-charges}
\begin{tabular}{|c|c|c|c|c|c|c|c|c|}
  \hline
  $s_{1}$ & $s_{2}$ & $s_{3}$ & $d_{1}$ & $d_{2}$ & $d_{3}$ & $u_{1}$ & $u_{2}$ & $u_{3}$ \\
  \hline
  0 & 0 & - & 0 & 0 & - & + & + & 0 \\
  \hline
\end{tabular}
\end{equation}

Although the interaction of an external field with dense quark
matter has been investigated by several authors
\cite{alf-berg-raj-NPB-02, iida}, the effect of the penetrating
$\widetilde{B}$ field on the superconducting gap structure was not
in the scope of these studies. However, as shown in our previous
paper \cite{MCFL}, the $\widetilde{B}$ field can change the gap
structure and lead to a new superconducting phase. To understand how
this can occur notice that due to the coupling of the charged quarks
with the external $\widetilde{B}$ field, the color-flavor space is
augmented by the charge operator $\widetilde{\mathbf{Q}}$, and
consequently the order parameter of the CFL splits in new
independent pieces. In this paper we show how the color
superconducting pairing is modified in the presence of a magnetic
field.

Let us introduce the rotated-charge projectors

\begin{equation}
\Omega_{0}={\rm diag}(1,1,0,1,1,0,0,0,1) \ ,
 \label{neut-omeg}
\end{equation}

\begin{equation}
\Omega_{+}={\rm diag}(0,0,0,0,0,0,1,1,0) \ , \label{pos-omeg}
\end{equation}

\begin{equation}
\Omega_{-}={\rm diag}(0,0,1,0,0,1,0,0,0)  \ , \label{neg-omeg}
\end{equation}
satisfying

\begin{equation}
\Omega_{\eta }\Omega_{\eta ^{\prime }}=\delta _{\eta \eta ^{\prime
}}\Omega_{\eta },\qquad \eta ,\eta ^{\prime }=0,+,- \ .
\label{omeg-algeb}
\end{equation}

\begin{equation}
\Omega_{0}+\Omega_{+}+\Omega_{-}=1 \ ,\label{omeg-sum}
\end{equation}

In terms of $\Omega_{+}$ and $\Omega_{-}$ the rotated charge
operator takes the form

\begin{equation}
\widetilde{\mathbf{Q}}=\sum\limits_{\eta =0,\pm }\eta \Omega _{\eta
}=\Omega_{+}-\Omega_{-}  \ .
 \label{Q-omeg-relat}
\end{equation}
Using the projectors (\ref{neut-omeg})-(\ref{neg-omeg}) we can
express the column fermion field in the 9-dimensional color-flavor
representation
$(s_{1},s_{2},s_{3},d_{1},d_{2},d_{3},u_{1},u_{2},u_{3})$ as the sum
\begin{equation}
\psi =\psi _{(0)}+\psi _{(+)}+\psi _{(-)}  \ , \label{ferm-sum}
\end{equation}
where subindexes $(0)$-, ($+/-)$ indicate fields of zero, positive
and negative rotated charge respectively defined by

\begin{equation}
\psi _{(0)}=\Omega_{0}\psi  \ ,\qquad \psi _{(+)}=\Omega_{+}\psi \
,\qquad \psi _{(-)}=\Omega _{-}\psi \ . \label{fields-def}
\end{equation}

The interaction of massless quarks with the external
rotated-magnetic field is described by the Lagrangian density term

\begin{equation}
L_{quarks}^{em}=\overline{\psi }(i\mathbf{\Pi}_{\mu }\gamma ^{\mu
})\psi \ , \label{quark-free-lag}
\end{equation}
with

\begin{equation}
\mathbf{\Pi} _{\mu }=i\partial _{\mu
}+\tilde{e}\widetilde{\mathbf{Q}}\widetilde{A}_{\mu}  \ .
\label{pi-operator}
\end{equation}

Given that the quarks have different rotated charges, the Lagrangian
density (\ref{quark-free-lag}) naturally splits in three terms

\begin{equation}
L_{quarks}^{em}=\overline{\psi} _{(0)} (i\partial _{\mu }\gamma
^{\mu })\psi _{(0)}+\overline{\psi} _{(+)}(i\partial _{\mu
}+\tilde{e}\widetilde{A}_{\mu})\gamma ^{\mu }\psi _{(+)}
+\overline{\psi }_{(-)}(i\partial _{\mu
}-\tilde{e}\widetilde{A}_{\mu})\gamma ^{\mu }\psi _{(-)} \ .
\label{qfree-lag-split}
\end{equation}

There is no reason to expect that the color-flavor structure of the
condensate will be the same as the CFL one. Instead, one would think
that the charge-separation of the fields enforced by the external
magnetic field should be somehow reflected in the color-condensate too.
Below we will show that this is indeed the case.

At this point we need to define a working model where we can
investigate the effects of the external magnetic field in a system
of massless quarks of three different flavors. With this aim, all
our calculations will be done in the context of a Nambu-Jona-Lasinio
(NJL) four-fermion interaction model, abstracted from one-gluon
exchange  interactions in QCD \cite{alf-raj-wil-99/537}. Although
this simple model disregards the effect of the $\widetilde
{B}$-field on the gluon dynamics and assumes the same NJL couplings
for the system with and without magnetic field, it keeps the main
attributes of the theory, providing the correct qualitative physics.

In order to find a reasonable ansatz for the gap in the presence of
a magnetic field, we will start considering the system without
magnetic field and will find the most general structure allowed for
the color superconducting (CS) condensate. The corresponding general
structure in the presence of the magnetic field will be a particular
case of the one at zero field, because the two structures have to
coincide in the limit of zero field and because the field reduces
the symmetry of the original theory to a subgroup of the original
group. We underline that such a general structure is found using
only very general symmetry arguments. The gap that minimizes the
free energy is just one of many possible particular condensate
patterns contained within the general structure: \textit{precisely
the one that retains the highest degree of symmetry}. This was the
guidance principle that led to the CFL ansatz
\cite{alf-raj-wil-99/537} in the case without magnetic field, and it
will be the same principle that will ultimately take us, when a
magnetic field is present, to a new ansatz: the MCFL one
\cite{MCFL}. As seen below, similarly to CFL, in the MCFL phase
quarks of all three colors and all three flavors pair. The main
difference is that the condensate distinguishes between gap
parameters that can get contributions from rotated charged quarks
and those that only get contributions from neutral quarks.

Let us recall that the gap matrix is formed by diquark condensates
$\langle q^{T}\mathcal{O}q\rangle$, where $q^{T}$ is the transposed
field operator, and
$\mathcal{O}=\mathcal{O}_{Dirac}\bigotimes\mathcal{O}_{flavor}\bigotimes\mathcal{O}_{color}$
is an operator acting through a direct product of the Dirac, flavor
and color spaces. 
Fermi statistics constraints $\mathcal{O}$ to be totally
antisymmetric, i.e. $\mathcal{O}^T=-\mathcal{O}$.

We are interested in a spin-zero condensate and will restrict our
attention to the spin-zero channel that is antisymmetric in Dirac
indices. Hence, the color-flavor structure of the gap should be
symmetric under simultaneous exchange of both color and flavor
indices. The Dirac structure is then given as $C\gamma_5$, where $C=
i \gamma_2 \gamma_0$ is the matrix of charge conjugation.

The one-gluon exchange interactions are attractive in the channels
that are antisymmetric under the exchange of color. This fact,
combined with the constraint imposed by the Fermi statistics
implies that the dominant contribution to the gap comes from terms
which are antisymmetric under the exchange of color indices and
antisymmetric under the exchange of flavor indices. This is the
color-antitriplet, flavor-antitriplet condensate. It can be shown
\cite{rajagop-schmitt-ph/0512043} that the antisymmetric gap can
be specified in general by only three gap parameters. Color
symmetric channels are repulsive, hence subdominant, but once the
color-antitriplet, flavor-antitriplet condensate is generated, a
color sextet, flavor sextet condensate is also induced due to
mixing \cite{Pisarski-rischke83-1999}. Following a derivation
similar to the one carried out in
Ref.~\cite{rajagop-schmitt-ph/0512043}, one can show that the
symmetric condensate can be specified in general by only six gap
parameters.

Therefore, the condensate structure can be written in general as

\begin{equation}
\mathbf{\Delta}=\Delta_{S_{1}} \mathbf{S}_1+\Delta_{S_{2}}
\mathbf{S}_2+\Delta_{S_{3}} \mathbf{S}_3+\Delta_{S_{4}}
\mathbf{S}_4+\Delta_{S_{5}} \mathbf{S}_5+\Delta_{S_{6}}
\mathbf{S}_6+\Delta_{A_{1}} \mathbf{A}_1+\Delta_{A_{2}}
\mathbf{A}_2+\Delta_{A_{3}} \mathbf{A}_3 \ , \label{general-gap}
\end{equation}
where the six symmetric and three antisymmetric independent
color-flavor structures are given by
\[
\mathbf{S}_1 =2\delta^{a1}\delta^{b1}\delta^{i1}\delta^{j1}\qquad
\mathbf{S}_2 =2\delta^{a2}\delta^{b2}\delta^{i2}\delta^{j2}\qquad
\mathbf{S}_3 =2\delta^{a3}\delta^{b3}\delta^{i3}\delta^{j3}\qquad
\mathbf{S}_4=(\delta^{a1}\delta^{b2}+\delta^{a2}\delta^{b1})(\delta^{i1}\delta^{j2}+\delta^{i2}\delta^{j1})\qquad
\]
\[
\mathbf{S}_5
=(\delta^{a1}\delta^{b3}+\delta^{a3}\delta^{b1})(\delta^{i1}\delta^{j3}+\delta^{i3}\delta^{j1})\qquad
\mathbf{S}_6
=(\delta^{a2}\delta^{b3}+\delta^{a3}\delta^{b2})(\delta^{i2}\delta^{j3}+\delta^{i3}\delta^{j2})
\]
\begin{equation}
\mathbf{A}_1 =\varepsilon_{ab1}\varepsilon_{ij1}\qquad \mathbf{A}_2
=\varepsilon_{ab2}\varepsilon_{ij2}\qquad \mathbf{A}_3
=\varepsilon_{ab3}\varepsilon_{ij3}\label{structure-elements}
\end{equation}
where $a,b$ denote color indices and $i,j$  denote flavor indices.
In the case of the NJL theory, since the quark interaction is
modelled by a point-like four-fermion term, the nine coefficients
$\Delta$ in Eq. (\ref{general-gap}) are independent of the momentum.

As mentioned above, the energetically favored condensate should be
the one retaining the highest degree of symmetry possible. In the
absence of a magnetic field, that corresponds to the so-called
color-flavor locked group $SU(3)_{C+L+R}$. Therefore, the original
$SU(3)_{C}\times SU(3)_{L}\times SU(3)_{R} \times U(1)_{B} \times
U(1)_{e.m.}$ symmetry is broken by the color condensate to the
diagonal subgroup $SU(3)_{C+L+R}$. This is the well-known CFL phase,
which results from Eq. (\ref{general-gap}) if one demands that
$\Delta_{S_1}=...=\Delta_{S_6}=\Delta_{S}$ and
$\Delta_{A_1}=...=\Delta_{A_3}=\Delta_{A}$. The CFL structure
parameter can be expressed as
$\mathbf{\Delta}_{CFL}=\Delta_{S}(\mathbf{U}+\mathbf{U}_{0})+\Delta_{A}(\mathbf{U}-\mathbf{U}_{0})$
with the color-flavor matrices $\mathbf{U}_{0}$ and $\mathbf{U}$
defined by

\begin{equation}
\label{u-zero-matrix}
 \mathbf{U}_{0}=\frac{1}{2}\sum_{i=1}^{6}\mathbf{S}_{i}+\frac{1}{2}\sum_{i=1}^{3}\mathbf{A}_{i} \ ,
\end{equation}

\begin{equation}
\label{u-matrix}
 \mathbf{U}=\frac{1}{2}\sum_{i=1}^{6}\mathbf{S}_{i}-\frac{1}{2}\sum_{i=1}^{3}\mathbf{A}_{i} \ .
\end{equation}

Let us consider now the situation with magnetic field. A
background magnetic field explicitly breaks the flavor symmetry of
the original theory to $SU(3)_{{\rm C}}\times SU(2)_{L}\times
SU(2)_{R} \times U(1)_B\times U(1)^{(-)}_A\times U(1)_{e.m.}$,
since u quarks have different electric charge than d and s quarks.
$U(1)^{(-)}_A$ is related to the existence of an anomaly-free
current given by a linear combination of the $s$, $d$, and $u$
axial currents \cite{miransky-shovkovy-02}. The subgroup that
ensures preserving the largest degree of symmetry in the CS gap is
$SU(2)_{C+L+R}$ in this case. Hence, the symmetry breaking induced
by the color condensate in the presence of a magnetic field is:
$SU(3)_{\rm C} \times SU(2)_L \times SU(2)_R \times U(1)_B \times
U^{(-)}(1)_A \times U(1)_{\rm e.m.} \rightarrow SU(2)_{{\rm
C}+L+R} \times {\widetilde {U}(1)}_{\rm e.m.}$. Notice that this
symmetry breaking preserves the rotated electromagnetism
${\widetilde {U}(1)}_{\rm e.m.}$.

The $SU(2)_{C+L+R}$ symmetry requires the invariance of the gap
under simultaneous flavor ($1 \leftrightarrow 2$) and color ($1
\leftrightarrow 2$) exchanges. This implies the following
equalities among the gap parameters in Eq.~(\ref{general-gap})

\begin{equation}
\Delta_{S'}\equiv\Delta_{S_1}=\Delta_{S_2} \ ,\qquad\
\Delta_{S}^{B}\equiv\Delta_{S_5}=\Delta_{S_6} \ ,\qquad\
\Delta_{A}^{B}\equiv\Delta_{A_1}=\Delta_{A_2} \
,\qquad\label{coeff=mcfl}
\end{equation}

Based on the above considerations, the gap structure in the presence
of a magnetic $\widetilde{B}$ field takes the form
\begin{equation}
\Delta=\left(
\begin{array}{ccccccccc}
2\Delta_{S'} & 0 & 0 & 0 & \Delta_{A}+\Delta_{S} & 0 & 0 & 0 & \Delta^{B}_{A}+\Delta^{B}_{S}\\
0 & 0 & 0 & \Delta_{S}-\Delta_{A} & 0 & 0 & 0 & 0& 0 \\
0 & 0 & 0 & 0 & 0 & 0 & \Delta^{B}_{S}-\Delta^{B}_{A} & 0 & 0 \\
0 & \Delta_{S}-\Delta_{A} & 0 & 0 & 0 & 0 & 0 & 0 & 0 \\
\Delta_{A}+\Delta_{S} & 0 & 0 & 0 & 2\Delta_{S'} & 0 & 0 & 0 &  \Delta^{B}_{A}+\Delta^{B}_{S}\\
0 & 0 & 0 & 0 & 0 & 0& 0 & \Delta^{B}_{S}-\Delta^{B}_{A} & 0 \\
0 & 0 & \Delta^{B}_{S}-\Delta^{B}_{A} & 0 & 0 & 0 & 0 & 0 & 0 \\
0 & 0 & 0 & 0 & 0 & \Delta^{B}_{S}-\Delta^{B}_{A} & 0 & 0&0\\
\Delta^{B}_{A}+\Delta^{B}_{S} & 0 & 0 & 0 &
\Delta^{B}_{A}+\Delta^{B}_{S} & 0 & 0 &0&2\Delta_{S''}
\label{order-parameter}
\end{array} \right)
\end{equation}

This is the so called MCFL ansatz. In going from (\ref{general-gap})
to (\ref{order-parameter}) we introduced the more convenient
notation $\Delta_{S''}\equiv \Delta_{S_{3}}$, $\Delta_{S}\equiv
\Delta_{S_{4}}$, $\Delta_{A}\equiv \Delta_{A_{3}}$. At weak magnetic
field, $\widetilde{B}$, the CFL and MCFL phases should be
continuously connected. In the limit of zero field the gap
parameters ought to satisfy $\Delta^{B}_{A}=\Delta_{A}$ and
$\Delta^{B}_{S}=\Delta_{S}=\Delta_{S'}=\Delta_{S"}$.

To trace back the physical origin of the different gap parameters
appearing in the MCFL structure (\ref{order-parameter}) we can see
that despite the overall $\widetilde{Q}$-neutrality of all the
pairs, they can be formed either by a pair of neutral or by a pair
of rotated charged quarks. Any pair formed by
$\widetilde{Q}$-charged quarks feel the background field directly
through the minimal coupling of the quarks in the pair with
$\widetilde{B}$. The gap parameters that get contributions from
pairs formed by $\widetilde{Q}$-charged quarks must be different
from the gap parameters getting contributions \textit{only} from
pairs of neutral quarks. The gaps $\Delta^B_{A/S}$ are
antisymmetric/symmetric combinations of condensates that get
contributions from both charged and neutral pairs. On the other
hand, the gaps $\Delta_{A/S,S',S"}$, are antisymmetric/symmetric
combinations of condensates that only get contributions from pairs
of neutral quarks. The only way the field can affect
$\Delta_{A/S,S',S"}$ is indirectly, through the system of highly
non-linear coupled gap equations.

The MCFL order parameter (\ref{order-parameter}) can be rewritten
in a compact way as

\begin{eqnarray}
\label{delta} \nonumber
 \mathbf{\Delta} &= &\Delta_{S'}(\mathbf{S}_1 +
 \mathbf{S}_2)+\Delta_{S"}\mathbf{S}_3+\frac{1}{2}[\Delta_{A}+\Delta_{S}]
 (\mathbf{S}_{4}+\mathbf{A}_3)+\frac{1}{2}[\Delta_{S}-\Delta_{A}](\mathbf{S}_{4}-\mathbf{A}_3)+
\\
 & + &\frac{1}{2}[\Delta_{A}^B+\Delta_{S}^B](\mathbf{S}_{5}+\mathbf{S}_{6}+\mathbf{A}_1
 +\mathbf{A}_2)+\frac{1}{2}[\Delta_{S}^B-\Delta_{A}^B](\mathbf{S}_{5}+\mathbf{S}_{6}-\mathbf{A}_1-\mathbf{A}_2)  \ .
\end{eqnarray}

Out of the four independent, symmetric gaps, all of which are due to
subleading interactions, one can single out the set of three gaps
$\Delta^{'}_{S}$, $\Delta^{''}_{S}$ and $\Delta_{S}$ that get
contributions only from pairs of neutral quarks. There is no reason
to expect that these three symmetric gaps will differ much in
magnitude. Hence, in a first approach to the problem, we will solve
the gap equations assuming $\Delta_S\simeq \Delta^{'}_{S} \simeq
\Delta^{''}_{S}$.

\section{Effective action in the presence of a rotated magnetic field
}\label{effec-act}

In this Section we introduce the mean-field action in the presence
of a constant rotated magnetic field in both coordinate and
momentum spaces. In our derivations we will follow a parallel
approach to the one developed for the case of zero field in Refs.
\cite{Bailin-Love,Pisarski-rischke83-1999}.

As discussed in the previous section, the rotated magnetic field
naturally separates the quark fields according to their
$\tilde{Q}$ charge. The $\widetilde{B}$-dependent mean-field
effective action is

\begin{eqnarray}
\label{b-coord-action} \nonumber I_{B}(\overline{\psi},\psi ) & = &
\int d^4xd^4y\Big\{ \frac{1}{2}\sum_{\tilde{Q}=0,\pm}
\Big[\overline{\psi }
_{(\tilde{Q})}(x)[G_{(\tilde{Q})0}^{+}]^{-1}(x,y)\psi
_{(\tilde{Q})}(y)+ \overline{\psi }
_{(\tilde{Q})C}(x)[G_{(-\tilde{Q})0}^{-}]^{-1}(x,y)
\psi_{(\tilde{Q})C}(y) \Big]
\\
&+ &\frac{1}{2}[\overline{\psi }_{(0)C}(x)\Delta ^{+}_{(0)}\psi
_{(0)}(y)+\overline{\psi }_{(+)C}(x)\Delta ^{+}_{(-)}\psi
_{(-)}(y)+\overline{\psi }_{(-)C}(x)\Delta ^{+}_{(+)}\psi
_{(+)}(y)+h.c.]\Big\} \ ,
\end{eqnarray}
where $\psi_{C}(x)= C \overline\psi^T(x)$ is the charge-conjugate spinor of $\psi(x)$.

In (\ref{b-coord-action}) symbols in parentheses indicate the
correspondence to neutral $(0)$, positively $(+)$ or negatively
$(-)$ $\widetilde{Q}$-charged fermions. Propagators for fields and
conjugated fields will be denoted with the customary $\pm$
supra-indexes. Explicit expressions of the inverse propagators are

\begin{equation}
\lbrack G_{(0)0}^{\pm}]^{-1}(x,y)=[i\gamma ^{\mu }\partial _{\mu
}\pm \mu \gamma ^{0}]\delta ^{4}(x-y) \ ,  \label{neut-x-inv-prop}
\end{equation}

\begin{equation}
\lbrack G_{(+)0}^{\pm }]^{-1}(x,y)=[i\gamma ^{\mu }\Pi ^{(+)}_{\mu
}\pm \mu \gamma ^{0}]\delta ^{4}(x-y) \ ,  \label{B-x-inv-prop}
\end{equation}

\begin{equation}
\lbrack G_{(-)0}^{\pm }]^{-1}(x,y)=[i\gamma ^{\mu }\Pi ^{(-)}_{\mu
}\pm \mu \gamma ^{0}]\delta ^{4}(x-y) \ ,  \label{B-x-inv-prop-ive}
\end{equation}

with

\begin{equation}
\Pi ^{(\pm)}_{\mu }=i\partial _{\mu
}\pm\widetilde{e}\widetilde{A}_{\mu } \ , \label{piplusminus}
\end{equation}

and the gap matrices are given by

\begin{equation}
\Delta _{(0)}^{+}=[\Delta_{S}+\Delta_{A}]^{+}
(\mathbf{U}_{0}-\mathbf{N}) \Omega
_{0}+[\Delta_{S}-\Delta_{A}]^{+} \mathbf{U} \Omega
_{0}+[\Delta^{B} _{S}+\Delta^{B} _{A}]^{+} \mathbf{N} \Omega _{0},
\label{101}
\end{equation}

\begin{equation}
\Delta _{(+)}^{+}=[\Delta^{B} _{S}-\Delta^{B}
_{A}]^{(+)}\mathbf{U} \Omega_{+},
 \label{102}
\end{equation}

\begin{equation}
\Delta _{(-)}^{+}=[\Delta^{B} _{S}-\Delta^{B}
_{A}]^{(+)}\mathbf{U}\Omega_{-}, \label{103}
\end{equation}
with the color-flavor matrix $\mathbf{N}$ defined as:
$\mathbf{N}=\frac{1}{2}
[\mathbf{S}_{5}+\mathbf{S}_{6}+\mathbf{A}_{1}+\mathbf{A}_{2}]$ .

\subsection{Effective action at $\widetilde{B}\neq0$ in momentum space}

The computation of the field-dependent quark propagators in momentum
space can be managed with the use of a method, originally developed
for charged fermions by Ritus \cite{Ritus:1978cj} and later on
extended to charged vector fields by Elizalde, Ferrer and Incera
\cite{efi-ext}. In Ritus' approach the diagonalization in momentum
space of the Green's functions of charged fermions in the presence
of a background magnetic field is carried out with the help of the
eigenfunction matrices $E_p(x)$. They are the wave functions of the
asymptotic states of charged fermions in a uniform magnetic field
and play the role in the magnetized medium of the usual plane-wave
(Fourier) functions $e^{i px}$ at zero field. This method has also
been used in the context of chiral mass generation in a magnetic
field \cite{orthonormality, ayala-et-al-ph/0606209}. Using the
$E_p(x)$ functions, we transform the propagators
(\ref{B-x-inv-prop})-(\ref{B-x-inv-prop-ive}) to momentum space
adequately. Below we describe the basic properties of the
transformation.

The transformation functions $E^{(\pm)}_{p}(x)$ for positively
($+$), and negatively ($-$) charged fermion fields are obtained as
the solutions of the field-dependent eigenvalue equation

\begin{equation}
(\Pi^{(\pm)}\cdot\gamma)E^{(\pm)}_{p}(x)=E^{(\pm)}_{p}(x)(\gamma\cdot\overline{p}^{(\pm)})
\ , \label{eigenproblem}
\end{equation}
with $\overline{p}^{(\pm )}$ given by

\begin{equation}  \label{pbar+}
\overline{p}^{(\pm )}=(p_{0},0,\pm
\sqrt{2|\widetilde{e}\widetilde{B}|k},p_{3}) \ ,
\end{equation}
and
\begin{equation}
E^{(\pm)}_{p}(x)=\sum\limits_{\sigma }E^{(\pm)}_{p\sigma
}(x)\Delta (\sigma ) \ , \label{9}
\end{equation}
with eigenfunctions

\begin{equation}
{E}^{(\pm)}_{p\sigma }(x)=\mathcal{N}%
_{n_{(\pm)}}e^{-i(p_{0}x^{0}+p_{2}x^{2}+p_{3}x^{3})}D_{n_{(\pm)}}(\varrho
_{(\pm)}) \ ,
 \label{Epsigma+}
\end{equation}
where $D_{n_{(\pm)}}(\varrho _{(\pm)})$ are the parabolic cylinder
functions with argument $\varrho _{(\pm)}$ defined by
\begin{equation}
\varrho _{(\pm)}=\sqrt{2|\widetilde{e}\widetilde{B}|}(x_{1}\pm p_{2}/\widetilde{e}%
\widetilde{B}) \ ,
 \label{rho+}
\end{equation}
and index $n_{(\pm)}$ given by
\begin{equation}  \label{normaliz-const+}
n_{(\pm)}\equiv n_{(\pm)}(k,\sigma)= k \pm \frac{\widetilde{e}\widetilde{B}}{2|%
\widetilde{e}\widetilde{B}|}\sigma-\frac{1}{2} \ , \qquad\qquad
n_{(\pm)}=0,1,2,...
\end{equation}
$k=0,1,2,3,...$ is the Landau level, and $\sigma$ is the spin
projection that can take values $\pm 1$. The normalization constant $%
\mathcal{N}_{n_{(\pm)}}$ is
\begin{equation}  \label{normaliz-const}
\mathcal{N}_{n_{(\pm)}}=(4\pi| \widetilde{e}\widetilde{B}|)^{\frac{1}{4}}/%
\sqrt{n_{(\pm)}!} \ .
\end{equation}

The spin matrices $\Delta (\sigma )$ in (\ref{9}), not to be
confused with the gap coefficients, are spin projectors. They are
defined as

\begin{equation}
\Delta (\sigma )= {\rm diag}(\delta _{\sigma 1},\delta _{\sigma
-1},\delta _{\sigma 1},\delta _{\sigma -1}),\qquad \sigma =\pm 1 \ ,
\label{10}
\end{equation}
and satisfy the following relations

\begin{equation}
\Delta \left( \pm \right) ^{\dagger }=\Delta \left( \pm \right) \
,\qquad \Delta(+) + \Delta (-)=1\ , \qquad \Delta \left( \pm
\right) \Delta \left( \pm \right) =\Delta \left( \pm \right)  \
,\qquad \Delta \left( \pm \right) \Delta \left( \mp \right) =0 \ ,
\end{equation}

\begin{equation}
\gamma ^{\shortparallel }\Delta \left( \pm \right) =\Delta \left(
\pm \right) \gamma ^{\shortparallel },\quad \gamma ^{\bot }\Delta
\left( \pm \right) =\Delta \left( \mp \right) \gamma ^{\bot } \ .
\label{24}
\end{equation}

In Eq. (\ref{24}) the notation $\gamma ^{\shortparallel }=(\gamma
^{0},\gamma ^{3})$ and $\gamma ^{\bot }=(\gamma ^{1},\gamma ^{2})$
was used.

Under the $E_p(x)$ functions, positively ($\psi _{(+)}$) and
negatively ($\psi _{(-)}$) charged fields transform according to
\begin{equation}
\psi
_{(\pm)}(x)=\sum\hspace{-0.5cm}\int\frac{d^{4}p}{(2\pi)^{4}}E_{p}^{(\pm)}(x)\psi
_{(\pm)}(p) \ ,\label{psi+-transf}
\end{equation}
\begin{equation}
\overline{\psi }_{(\pm)}(x)=\sum\hspace{-0.5cm}\int\frac{d^{4}p}{(2\pi)^{4}}
\overline{\psi }%
_{(\pm)}(p)\overline{E}_{p}^{(\pm)}(x)  \ , \label{psibar+-transf}
\end{equation}
where $\overline{E}_{p}^{(\pm)}(x)=\gamma
_{0}({E}_{p}^{(\pm)}(x))^{\dag }\gamma _{0}$ and
$\sum\hspace{-0.35cm}\int\frac{d^{4}p}{(2\pi)^{4}}\equiv
\sum_{k=0}^{\infty}\int \frac{dp_{0}dp_{2}dp_{3}}{(2\pi)^{4}}$.

One can show that
\begin{equation}  \label{gamma-piplus-propert}
[\gamma_{\mu}(\Pi_{(+)\mu}\pm \mu\delta_{\mu0})]{E}^{(+)}_{p}(x)={E}%
^{(+)}_{p}(x)[\gamma_{\mu}(\overline{p}^{(+)}_{\mu}\pm
\mu\delta_{\mu0})] \ ,
\end{equation}
and
\begin{equation}  \label{gamma-piminus-propert}
[\gamma_{\mu}(\Pi_{(-)\mu}\pm \mu\delta_{\mu0})]{E}^{(-)}_{p}(x)={E}%
^{(-)}_{p}(x)[\gamma_{\mu}(\overline{p}^{(-)}_{\mu}\pm
\mu\delta_{\mu0})] \ .
\end{equation}

Since the charge conjugate of a positively (negatively) charged
field is a negatively (positively) charged field, the charge
conjugate fields transform as

\begin{equation}  \label{conjpsiplustransf}
\psi_{(+)C}(x)=\sum\hspace{-0.5cm}\int\frac{d^{4}p}{(2\pi)^{4}}E^{(-)}_{p}(x)\psi_{(+)C}(p),
\end{equation}
\begin{equation}  \label{conjpsiminustransf}
\psi_{(-)C}(x)=\sum\hspace{-0.5cm}\int\frac{d^{4}p}{(2\pi)^{4}}E^{(+)}_{p}(x)\psi_{(-)C}(p)
\ .
\end{equation}

This transformation dictates what fields should form the
components of positively and negatively charged Nambu-Gorkov
fields.

In terms of Nambu-Gorkov fields $\Psi$, the effective action in
the presence of magnetic field $\widetilde{B}$ takes the form

\begin{eqnarray}  \label{b-action}
I^{B}(\overline{\psi},\psi)=\int d^4x \,d^4 y \,\bar \Psi(x) {\cal
S}^{-1}(x,y) \Psi(y) \qquad\qquad\qquad\qquad\qquad \nonumber
\\
=\frac{1}{2}\int\frac{d^{4}p}{(2\pi)^{4}} \overline{\Psi}%
_{0}(p){\cal
S}^{-1}_{0}(p)\Psi_{0}(p)+\frac{1}{2}\sum\hspace{-0.5cm}\int\frac{d^{4}p}{(2\pi)^{4}}
\overline{\Psi}_{+}(p){\cal
S}^{-1}_{+}(p)\Psi_{+}(p)+\frac{1}{2}\sum\hspace{-0.5cm}\int\frac{d^{4}p}{(2\pi)^{4}}
\overline{\Psi}_{-}(p){\cal S}^{-1}_{-}(p)\Psi_{-}(p) \ ,
\end{eqnarray}
where
\begin{eqnarray}  \label{neutr-inv-propg}
{\cal S}^{-1}_{(0)}(p)=\left(
\begin{array}{cc}
[G_{(0)0}^{+}]^{-1}(p) & \Delta_{(0)}^{-} \\
&  \\
\Delta_{(0)}^{+} & [G_{(0)0}^{-}]^{-1}(p)
\end{array}
\right) \ ,
\end{eqnarray}

\begin{eqnarray}  \label{posit-inv-propg}
{\cal S}^{-1}_{(+)}(p)=\left(
\begin{array}{cc}
[G_{(+)0}^{+}]^{-1}(p) & \Delta_{(+)}^{-} \\
&  \\
\Delta_{(+)}^{+} & [G_{(+)0}^{-}]^{-1}(p)
\end{array}
\right) \ ,
\end{eqnarray}
\begin{eqnarray}  \label{negat-inv-propg}
{\cal S}^{-1}_{(-)}(p)=\left(
\begin{array}{cc}
[G_{(-)0}^{+}]^{-1}(p) & \Delta_{(-)}^{-} \\
&  \\
\Delta_{(-)}^{+} & [G_{(-)0}^{-}]^{-1}(p)
\end{array}
\right) \ ,
\end{eqnarray}
and
\begin{equation}
\Delta _{(0)}^{-}=\gamma _{0}[\Delta _{(0)}^{+}]^{\dagger }\gamma
_{0},\qquad \Delta _{(+)}^{-}=\gamma _{0}[\Delta
_{(+)}^{+}]^{\dagger }\gamma _{0},\qquad \Delta _{(-)}^{-}=\gamma
_{0}[\Delta _{(-)}^{+}]^{\dagger }\gamma _{0} \ .
\end{equation}

The Nambu-Gorkov fermion fields corresponding to the different
$\widetilde{Q}$ charges are defined by

\begin{equation}
\Psi _{0}=\left(
\begin{array}{c}
\psi _{(0)} \\
\psi _{(0)C}
\end{array}
\right) \ ,
\end{equation}
for the neutral field,
\begin{equation}
\Psi _{+}=\left(
\begin{array}{c}
\psi _{(+)} \\
\psi _{(-)C}
\end{array}
\right) \ ,
\end{equation}
for the positive field, and
\begin{equation}
\Psi _{-}=\left(
\begin{array}{c}
\psi _{(-)} \\
\psi _{(+)C}
\end{array}
\right) \ ,
\end{equation}
for the negative field.

As noticed before, the positive (negative) Nambu-Gorkov field is
formed by the positive (negative) fermion field and the charge
conjugate of the negative (positive) field. This means that, as it
should be, the rotated charge of the up and down components of a
given Nambu-Gorkov field are the same. This in turn determines the
nature of the $\tilde{Q}$-charges of the fields entering in a
given pair, which we know should have zero overall $\tilde{Q}$
charge.

The bare inverse propagator of the neutral field is

\begin{equation}
\lbrack G_{(0)0}^{\pm }]^{-1}(p)=\gamma _{\mu }(p_{\mu }\pm \mu
\delta _{\mu 0}) \ , \label{neut-bareprop+-}
\end{equation}
Here the momentum is the usual $p=(p_{0},p_{1},p_{2},p_{3})$ of the
case with no background field.

For the positively and negatively charged fields the bare inverse
propagators are

\begin{equation}
\lbrack G_{(+)0}^{\pm }]^{-1}(p)=\gamma _{\mu }(\overline{p}%
_{\mu }^{(+)}\pm \mu \delta _{\mu 0}) \ , \label{pos-bareprop+-}
\end{equation}
\begin{equation}
\lbrack G_{(-)0}^{\pm }]^{-1}(p)=\gamma _{\mu }(\overline{p}%
_{\mu }^{(-)}\pm \mu \delta _{\mu 0})  \ . \label{neg-bareprop+-}
\end{equation}
respectively.

\subsection{Nambu-Gorkov propagators in a color-flavor rotated basis}
\label{NG-propagators}

In order to find the propagators in Nambu-Gorkov space one has to
invert the matrices (\ref{neutr-inv-propg})-(\ref{negat-inv-propg}),
whose color-flavor structure is rather complicated. At this point we
find convenient to perform a rotation that can partially diagonalize
the components of the Nambu-Gorkov propagators in color-flavor
space.

Let us define a new basis for the fermion fields

\begin{equation}
 \label{CFL-basis}
 \psi_{ai} =\frac{1}{\sqrt{2}} \sum_{A=1}^9 \lambda_{ai}^A \psi^A \ , \qquad
\psi_{C,ai} = \frac{1}{\sqrt{2}} \sum_{A=1}^9 (\lambda_{ai}^A)^T
\psi^A \ ,
 \end{equation}
where the indices  $a$ and $i$ refer to color and flavor
respectively, and $\lambda^A$, for $A=1, \dots,8$ are the Gell-Mann
matrices, while $\lambda^9 = \sqrt{\frac 23} \mathbf{1}$. Given that
in the CFL case the Nambu-Gorkov quark propagators become
color-flavor diagonal \cite{son-stephanov-prd61-074012} in the basis
(\ref{CFL-basis}), we call (\ref{CFL-basis}) the CFL basis.

In the CFL basis the quark kinetic term (\ref{quark-free-lag})
 becomes
\begin{equation}
L_{quarks}^{em}=i \overline{\psi }^A \gamma^\mu \left(i \partial_\mu
\delta^{AB} - {\tilde e} \,{\mathbf{\widetilde Q}}^{AB} {\tilde
A}_\mu \right)\psi^B \ ,
\end{equation}
where the charge matrix is
\begin{equation}
{\mathbf{\widetilde Q}}^{AB} = - \frac 12 {\rm Tr} \left(
[\lambda^A, \lambda^B ] \mathbf{Q} \right) \ .
\end{equation}
In our conventions, $\mathbf{Q} = - \frac{\lambda_8}{\sqrt{3}}$, and
thus $\mathbf{{\widetilde Q}}^{AB}  = \frac{2 i}{\sqrt{3}} f^{AB8}$
is expressed in terms of the antisymmetric structure constants of
$SU(3)$. In the basis (\ref{CFL-basis}) one can as well find the
charge-projector operators
\begin{equation}
\Omega^{AB}_{0} = {\rm diag}(1,1,1,0,0,0,0,1,1) \ ,
\label{neut-omegCFL}
\end{equation}
and
\begin{equation}
\Omega^{AB}_{\pm} = \frac{1}{2}[(\delta^{A4} \delta^{B4} +
\delta^{A5} \delta^{B5} + \delta^{A6} \delta^{B6}+ \delta^{A7}
\delta^{B7} )\pm i(\delta^{A4} \delta^{B5} - \delta^{A5}
\delta^{B4}+\delta^{A6} \delta^{B7} - \delta^{A7} \delta^{B6})] \  ,
\label{pos-omegCFL}
\end{equation}
which obey the conditions (\ref{omeg-algeb})-(\ref{Q-omeg-relat}).
Furthermore, one has $(\Omega_{\pm})^T = \Omega_{\mp}$, so that
$\Omega_{\pm} \psi_C = (\psi_{\mp})_C$.

From Eq.~(\ref{CFL-basis}) one can deduce the transformation law of
the gap matrix, which now reads

\begin{equation}
 \Delta^{AB} = \frac 12 (\lambda^A)^T_{ai}
\Delta_{ab}^{ij} \lambda^B_{bj} \ .
\end{equation}
After taking into account the approximation $\Delta_S \approx
\Delta_{S'} \approx \Delta_{S"}$ in the gap matrix
(\ref{order-parameter}) one obtains the gap parameters in the new
basis to be
\begin{subequations}
\label{gapCFLbasis}
\begin{eqnarray}
\Delta^{11} & = & \Delta^{22} = \Delta^{33} =  \Delta_S -\Delta_A  \ , \\
\Delta^{44} & = & \Delta^{55} = \Delta^{66} = \Delta^{77} =  \Delta_S^B -\Delta_A^B \ , \\
\Delta^{88} & = & \frac 13 \left( \Delta_A -4 \Delta^B_A + 7 \Delta_S - 4 \Delta^B_S\right) \ , \\
\Delta^{89} & = & \Delta^{98}  =\frac{\sqrt{2}}{3} \left(\Delta_A -
\Delta^B_A + \Delta_S - \Delta^B_S
\right) \ , \\
\Delta^{99} & = & \frac 23 \left( \Delta_A +2 \Delta^B_A + 4
\Delta_S + 2 \Delta^B_S\right) \ ,
\end{eqnarray}
\end{subequations}
while all the remaining elements of $\Delta^{AB}$ are zero.

In the CFL basis the Nambu-Gorkov effective action (\ref{b-action})
can be written as

\begin{equation}  \label{CFL-Gorkov-action}
 I(\overline{\psi},{\psi})=
 \int \frac{d^{4}p}{(2\pi)^4}
 \bar \Psi^{A}_0(p) [{\cal S}^{-1}_{(0)}(p)]^{AB} \Psi^B_0(p)
+ \sum\hspace{-0.5cm}\int\frac{d^{4}p}{(2\pi)^{4}}\Big[\bar
\Psi^{A}_+(p) [{\cal S}^{-1}_{(+)}(p)]^{AB} \Psi^B_+(p) +
\bar\Psi^{A}_-(p) [{\cal S}^{-1}_{(-)}(p)]^{AB} \Psi^B_-(p) \Big] ,
\end{equation}

As follows we find convenient to  express all propagators in terms
of energy projectors. For completion, in Appendix \ref{App-A} we
present the details on how they are generalized for charged
particles in the presence of an external magnetic field.

For the charged fields the Nambu-Gorkov matrix reads
\begin{equation}
{\cal S}^{AB}_{(\pm)} (p)= \delta^{AB} \,\left(
\begin{array}{cc}
S_{(\pm)_{11}}(\overline{p}^{(\pm )}) & S_{(\pm)_{12}}  (\overline{p}^{(\pm )})\\
S_{(\pm)_{21}} (\overline{p}^{(\pm )}) & S_{(\pm)_{22}}
(\overline{p}^{(\pm )})
\end{array}
\right) \ ,
\end{equation}
where the indexes $A,B$ take values $4,5,6,7$ only and the
subindices $11, 12, 21, 22$ specify the entry of the Nambu-Gorkov
matrix. Notice that ${\cal S}^{AB}_{(\pm)} (p)$ depends on the
Minkowski momenta $\overline{p}^{(\pm )}= (p_0, {\bf \bar
p}^{(\pm)})$ defined in Eq.~(\ref{pbar+}). To avoid confusion, we
will always denote with caligraphic letters the Nambu-Gorkov
matrixes, and with capital letters the elements of the matrixes.

For massless quarks, the elements of the Nambu-Gorkov matrix for
the positive charged quarks are
\begin{subequations}
\begin{eqnarray}
 S_{(+)_{11/22}} (\overline{p}^{(+)}) & = &\frac{{\tilde \Lambda}_{(+)}^{\pm}(\overline{p}) \gamma^0
 \left(p_0 \mp \mu \pm  |{\bf {\bar p^{(+)}}}| \right)}
{p_0^2 - (\mu -  |{\bf {\bar p^{(+)}}}|)^2 - (\Delta_{A}^B
-\Delta^B_S)^2}  + \frac{{\tilde \Lambda}_{(+)}^{\mp}(\overline{p}) \gamma^0
\left(p_0 \mp \mu \mp  |{\bf {\bar p^{(+)}}}| \right)} {p_0^2 - (\mu
+
|{\bf {\bar p^{(+)}}}|)^2 - ({\Delta}_{A}^B -{ \Delta}_S^B)^2}  \ , \\
S_{(+)_{21}} (\overline{p}^{(+)}) & = & \gamma_5 \left\{
\frac{{\tilde \Lambda}_{(+)}^- (\overline{p})\left(\Delta_S^B - \Delta_A^B \right)
} {p_0^2 - (\mu -  |{\bf {\bar p^{(+)}}}|)^2 - (\Delta_{A}^B
-\Delta^B_S)^2} + \frac{{\tilde \Lambda}_{(+)}^+ (\overline{p})\left({ \Delta}_S^B
- { \Delta}_A^B \right) } {p_0^2 - (\mu + |{\bf {\bar p^{(+)}}}|)^2
- ({ \Delta}_{A}^B -{ \Delta}_S^B)^2}
\right\} \ , \\
S_{(+)_{12}} (\overline{p}^{(+)}) & = & - \gamma_5 \left\{
\frac{{\tilde \Lambda}_{(+)}^+ (\overline{p})\left(\Delta_S^B - \Delta_A^B
\right)^*} {p_0^2 - (\mu -  |{\bf {\bar p^{(+)}}}|)^2 -
(\Delta_{A}^B -\Delta^B_S)^2} + \frac{{\tilde \Lambda}_{(+)}^-(\overline{p})
\left({ \Delta}_S^B - { \Delta}_A^B \right)^*} {p_0^2 - (\mu + |{\bf
{\bar p^{(+)}}}|)^2 - ({ \Delta}_{A}^B -{\ \Delta}_S^B)^2} \right\}
\ ,
\end{eqnarray}
\end{subequations}%

Following the definition  of the momenta (\ref{pbar+}), we have
$|{\bf {\bar p^{(+)}}}| = \sqrt{ 2 | {\tilde e}{\tilde B}| k +
p^2_3}$. We use the symbol ${\tilde \Lambda}_{(+)}^{\pm}(\overline{p})$ to denote
positive/negative energy projectors of positively charged
quasiquarks in the presence of the external field, as defined in Appendix \ref{App-A}.

The propagator for the negatively charged fields keeps the same
structure, with the only change $\overline{p}^{(+)} \rightarrow
\overline{p}^{(-)}$. Its off-diagonal term satisfies the
equation,

\begin{equation}
 S_{(-)_{21}} (x,x)=S_{(+)_{21}}(x,x) \ , \label{requir}
 \end{equation}
since

\bea \label{Sneg}
 S_{(\pm)_{21}} (x,x) & = &\sum\hspace{-0.5cm}\int\frac{d^{4}p}{(2\pi)^{4}}
E^\pm_{p}(x)\gamma_5 \left\{ \frac{{\tilde
\Lambda}_{(\pm)}^-(\overline{p}) \left(\Delta_S^B - \Delta_A^B
\right) } {p_0^2 - (\mu -  |{\bf {\bar
p^{(\pm)}}}|)^2 - (\Delta_{A}^B -\Delta^B_S)^2} \right. \nonumber \\
 & + & \left.
\frac{{\tilde \Lambda}_{(\pm)}^+(\overline{p}) \left({ \Delta}_S^B - { \Delta}_A^B
\right) } {p_0^2 - (\mu + |{\bf {\bar p^{(\pm)}}}|)^2 - ({
\Delta}_{A}^B -{\Delta}_S^B)^2} \right\} \ {\bar E}^\pm_{p}(x) \ ,
\eea
 so after integrating in $p_2$ and using the
relation

\begin{equation}
 \int^{\infty}_{-\infty}dp_{2}D_{n}(\rho_{(\pm)})D_{n'}(\rho_{(\pm)})=n!\sqrt{2\pi}\frac{\tilde e\tilde B}{\sqrt{2\tilde e\tilde
 B}}~\delta_{nn'}
 \end{equation}
we obtain
\bea \label{Sxx}
 S_{(\pm)_{21}} (x,x) & = & \frac{\tilde e\tilde
B}{2}\sum_{k=0}^{\infty}\int\frac{dp_{0}dp_{3}}{(2\pi)^3} (\Delta_S^B - \Delta_A^B
)\gamma_5 \left\{ \frac{1} {p_0^2 - (\mu -  |{\bf {\bar
p}}|)^2 - (\Delta_{A}^B -\Delta^B_S)^2}
 \right. \nonumber \\
 & + & \left.
\frac{1} {p_0^2 - (\mu + |{\bf {\bar p}}|)^2 - ({ \Delta}_{A}^B -{
\Delta}_S^B)^2} \right\} \ ,
\eea
where $|{\bf {\bar p}}|$ stands for
the common modulus of the three-momentum of positively and
negatively charged quasiparticles, since $|{\bf {\bar
p}}|\equiv|{\bf {\bar p^+}}| = |{\bf {\bar p^-}}|$.  From now on we
will use the short notation $|{\bf {\bar p}}|$ for the modulus the
three momentum of charged quasiparticles.

The propagator for the neutral fields in the CFL basis with indexes
$A,B=1,2,...9$ is given by

\begin{equation}
{\cal S}_{(0)}^{AB} (p) =\left(
\begin{array}{ccccccccc}
{\cal S}_{(0)}^{11}(p)  & 0 & 0 & 0 & 0 & 0 & 0 & 0 & 0 \\
0 & {\cal S}_{(0)}^{22} (p)  & 0 &0& 0 & 0 & 0 & 0& 0 \\
0 & 0 & {\cal S}_{(0)}^{33} (p) & 0 & 0 & 0 & 0 & 0 & 0 \\
0 & 0 & 0 & 0 & 0 & 0 & 0 & 0 & 0 \\
0& 0 & 0 & 0 & 0 & 0 & 0 & 0 &  0\\
0 & 0 & 0 & 0 & 0 & 0& 0 & 0 & 0 \\
0 & 0 & 0 & 0 & 0 & 0 & 0 & 0 & 0 \\
0 & 0 & 0 & 0 & 0 & 0 & 0 & {\cal S}_{(0)}^{88}(p)  &{\cal S}_{(0)}^{89}(p)  \\
0 & 0 & 0 & 0 &0 & 0 & 0 & {\cal S}_{(0)}^{98}(p)  & {\cal
S}_{(0)}^{99}(p) \label{neutralprop}
\end{array} \right) \ ,
\end{equation}
where each component is a $2 \times 2$ Nambu-Gorkov matrix
\begin{eqnarray}
{\cal S}_{(0)}^{AB} (p)   = \left(
\begin{array}{cc}
S_{(0)_{11}}^{AB}(p)& S_{(0)_{12}}^{AB}(p)\\
S_{(0)_{21}}^{AB}(p)& S_{(0)_{22}}^{AB}(p)
\end{array} \right)  \ .
\end{eqnarray}

One finds that ${\cal S}_{(0)}^{11} = {\cal S}_{(0)}^{22}= {\cal
S}_{(0)}^{33}$, with Nambu-Gorkov components given by
\begin{subequations}
\label{propaCFL}
\begin{eqnarray}
 S_{(0)_{11/22}}^{11}(p) & = &\frac{\Lambda^{\pm}(p) \gamma^0
 \left(p_0 \mp \mu \pm  |{\bf p}| \right)}
{p_0^2 - (\epsilon^0_p)^2}  + \frac{\Lambda^{\mp}(p) \gamma^0 \left(p_0
\mp \mu \mp  |{\bf p}| \right)}
{p_0^2 - ({\bar \epsilon^0}_p)^2}  \ , \\
S_{(0)_{21}}^{11}(p) & = & \gamma_5 \left\{ \frac{\Lambda^-(p)
\left(\Delta_S -\Delta_A \right)} {p_0^2 - (\epsilon^0_p)^2}  +
\frac{\Lambda^+(p) \left({\Delta}_S -{ \Delta}_A \right)} {p_0^2 -
({\bar
\epsilon}^0_p)^2} \right\} \ , \\
S_{(0)_{12}}^{11}(p) & = & - \gamma_5 \left\{ \frac{\Lambda^+(p)
\left(\Delta_S-\Delta_A \right)^*} {p_0^2 - (\epsilon^0_p)^2}  +
\frac{\Lambda^-(p) \left({ \Delta}_S -{ \Delta}_A \right)^*} {p_0^2 -
({\bar \epsilon}^0_p)^2} \right\}
 \ ,
\end{eqnarray}
\end{subequations}
with
\begin{equation}
\epsilon^0_p  =  \sqrt{(\mu -  |{\bf  p}|)^2 + (\Delta_{A}
-\Delta_S)^2} \ , \qquad {\bar \epsilon}^0_p  =  \sqrt{ (\mu + |{\bf
p}|)^2 + ({ \Delta}_{A} -{ \Delta}_S)^2 } \ .
 \end{equation}

The remaining block matrix for the CFL coefficients $A=8,9$ in
(\ref{neutralprop})
\begin{eqnarray}
\left(
\begin{array}{cc}
{\cal S}^{88}_{(0)} & {\cal S}^{89}_{(0)}\\
{\cal S}^{98}_{(0)}& {\cal S}^{99}_{(0)}
\end{array} \right) \
\label{matrixM}
\end{eqnarray}
has matrix elements which are $2 \times 2$ matrixes in the
Nambu-Gorkov space. Given that the symmetric gap $\Delta_{S}$ comes
from subdominant channels, we can assume, as in the CFL case, that
$\Delta_{S}\ll\Delta_{A}$. Using this approximation the Nambu-Gorkov
components of each of the matrix element of (\ref{matrixM}) can be
expressed as

\begin{subequations}
\label{89CFLB}
\begin{eqnarray}
 S^{AB}_{(0)_{11/22}} & = & \frac{\Lambda^{\pm}(p) \gamma^0
 \left(p_0 \mp \mu \pm  |{\bf p}| \right) A_{AB}}{\left(p^2_0 - (\epsilon_p^a)^2 \right)\left(p^2_0 - (\epsilon_p^b)^2 \right) }
  + \frac{\Lambda^{\mp}(p)\gamma^0
\left(p_0 \mp \mu \mp  |{\bf p}| \right) \bar A_{AB}}{\left(p^2_0 -
(\bar \epsilon_p^a)^2 \right)\left(p^2_0 - (\bar \epsilon_p^b)^2
\right)}
  \ , \\
S^{AB}_{(0)_{21}} & = &  \gamma_5 \left\{
\frac{\Lambda^-(p)\,B^{AB}}{\left(p^2_0 - (\epsilon_p^a)^2
\right)\left(p^2_0 - (\epsilon_p^b)^2 \right)} + \frac{\Lambda^+(p)
\,\bar B^{AB}}{\left(p^2_0 - (\bar \epsilon_p^a)^2
\right)\left(p^2_0 - (\bar
\epsilon_p^b)^2 \right) } \right\} \ , \\
S^{AB}_{(0)_{12}} & = & -\gamma_5 \left\{
\frac{\Lambda^+(p)\,(B^{AB})^*}{\left(p^2_0 - (\epsilon_p^a)^2
\right)\left(p^2_0 - (\epsilon_p^b)^2 \right)} + \frac{\Lambda^-(p) \,
(\bar B^{AB})^*}{\left(p^2_0 - (\bar \epsilon_p^a)^2
\right)\left(p^2_0 - (\bar \epsilon_p^b)^2 \right) } \right\} \ ,
\end{eqnarray}
 \end{subequations}%
where
\begin{equation}
\epsilon^{a/b}_p  =  \sqrt{(\mu -  |{\bf  p}|)^2 + (\Delta_{a/b})^2}
\ , \qquad {\bar \epsilon}^{a/b}_p  =  \sqrt{ (\mu +  |{\bf  p}|)^2
+ ({ \Delta}_{a/b})^2 } \ ,
 \end{equation}
with
\begin{equation}
\label{gaps-ab} \Delta^2_{a/b}  =  \frac 12 \left( 4 \Delta^2_N +
\Delta^2_A \pm \Delta_A \sqrt{\Delta^2_A + 8 \Delta^2_N } \right) \
.
\end{equation}
and we have defined for convenience
\begin{equation}
\Delta_N = \Delta_A^B + \Delta_S^B \ .
\end{equation}

The CFL submatrices $A$ and $B$  read
\begin{equation}
 A = \left(
\begin{array}{cc}
 r^2  - \frac 29 \left(\Delta_A - \Delta_N\right)^2 - \frac 49\left(\Delta_A + 2 \Delta_N \right)^2
 & \frac{\sqrt{2}}{3} \Delta_A  \left( \Delta_A -\Delta_N \right) \\
 \frac{\sqrt{2}}{3}  \Delta_A  \left(\Delta_A  - \Delta_N \right)  &
r^2   - \frac 29 \left(\Delta_A - \Delta_N\right)^2 - \frac
19\left(\Delta_A - 4 \Delta_N  \right)^2
\end{array}
\right)
 \ ,
\end{equation}

\begin{equation}
 B  = \left(
\begin{array}{cc}
 \frac{r^2}{3} \left(\Delta_A - 4\Delta_N\right) + \frac 43 \Delta_N^2\left(\Delta_A + 2 \Delta_N \right)
 & \frac{\sqrt{2}}{3}  \left( \Delta_A -\Delta_N \right) \left(r^2 - 2 \Delta_N^2 \right)\\
 \frac{\sqrt{2}}{3}  \left(\Delta_A  - \Delta_N \right) \left(r^2 - 2 \Delta_N^2 \right)  &
\frac{2r^2}{3}  \left(\Delta_A + 2\Delta_N\right) + \frac{2
\Delta_N^2}{3} \left(\Delta_A - 4 \Delta_N  \right)
\end{array}
\right)
 \ ,
\end{equation}
where we used the following shorthand
$ r^2 \equiv p^2_0 - ( |{\bf p}| - \mu)^2$. The matrices $\bar A$
and $\bar B$ are the corresponding antigap quantities. They are
obtained simply by replacing  $r^2 \rightarrow \bar {r}^2 \equiv
p^2_0 - ( |{\bf p}| + \mu)^2$.

\section{Gap Equations and Solutions in the presence of a rotated magnetic field
} \label{Sec-gap-eq}

The evaluation of the different fermionic gaps can be done by
writing the Schwinger-Dyson equations. In coordinate space the  QCD
gap equation reads
\begin{equation} \label{gap-eq} \Delta^+(x,y) = i\frac{g^2}{4} \lambda_A^T\,
\gamma^\mu\, S_{21}(x,y) \gamma^\nu\, \lambda_B D^{AB}_{\mu \nu}
(x,y) \ , \end{equation}
 where, for simplicity, we have
omitted explicit color and flavor indices in the gap and fermion
propagator. Here $D^{AB}_{\mu \nu}$ is the gluon propagator.

 We leave for a future project the
study of Eq.~(\ref{gap-eq}) in the high density/weak coupling limit
of QCD in the presence of a magnetic field. It should be noted,
though, that gluons can also feel the presence of a rotated magnetic
field because some of the gluons carry ${\widetilde Q}$-charges, and
thus they minimally couple to the external ${\widetilde B}$-field.
Debye and Meissner masses, as well as Landau damping, should then be
affected by ${\widetilde B}$. Hence, before studying the QCD gap
equation (\ref{gap-eq}) in the presence of a magnetic field, it is
necessary to study the behavior of the gluon propagator for
different energy scales, so as to properly assess its effect in the
gap equation.

In the NJL model abstracted from one-gluon exchange that we are
using in this paper the gap equation can be obtained from
Eq.~(\ref{gap-eq}) simply by substituting the gluon propagator by

\begin{equation}
D^{AB}_{\mu \nu} (x,y) = \frac{1}{\Lambda^2} \,g_{\mu \nu}\,
\delta^{AB} \,\delta^{(4)}(x -y) \ .
\end{equation}
and using the quark propagator in the presence of the rotated
magnetic field found in Section \ref{effec-act}.

The NJL model is characterized by a coupling constant $g$ and an
ultraviolet cutoff $\Lambda$. The ultraviolet cutoff should be
much larger than any of the typical energy scales of the system,
such as the chemical potential $\mu$ and the magnetic energies
$\sqrt{{\tilde e}{\tilde B} }$. In other studies of color
superconductivity within the NJL model the values of $g$ and
$\Lambda$ were chosen so to match some QCD vacuum properties, and
consequently hoping to get also the correct approximated
quantitative results for the gaps. We will follow the same
philosophy here, however, noticing again that this completely
ignores the effect of the magnetic field on the gluon dynamics.

\subsection{Gap equations for arbitrary values of the magnetic field}
\label{gen-gapeq}

Using the transformation rules given in Eq. (\ref{CFL-basis}) one
can re-write the NJL gap equation in the CFL basis

\begin{equation}
 \label{mother-gap}
(\Delta^+)^{{\bar B} C}  = -i  \frac{g^2 }{8 \Lambda^2} \sum_{B,
{\bar C} =1}^9 \left\{ {\rm Tr} \left(\lambda^{\bar B} \lambda^{\bar
C} \lambda^C \lambda^B \right) - \frac 13 {\rm Tr}
\left(\lambda^{\bar B} \lambda^{\bar C}\right){\rm Tr} \left(
\lambda^C \lambda^B \right) \right\} \gamma^\mu S_{({\widetilde
Q})_{21}}^{B {\bar C}}(x,x) \gamma^\mu \ .
\end{equation}

After computing the above $U(3)$ traces, taking into account the
particular CFL-basis structure of the propagator, where $S^{11}_{21}
=S^{22}_{21} =S^{33}_{21}$, and keeping in mind that the relation
(\ref{requir}) is held, we can write down more explicitly the gap
equations for every element of $\Delta^{AB}$. They read

\begin{subequations}
\begin{equation}
\label{gapeq11}
 (\Delta^+)^{11}  =  - i\frac{g^2 }{4 \Lambda^2}
\gamma^\mu\left\{ -\frac 53 S^{11}_{(0)_{21}} (x,x) + \frac 13
\left(S^{88}_{(0)_{21}}(x,x) + 2 \sqrt{2} S^{89}_{(0)_{21}}(x,x) + 2
S^{99}_{(0)_{21}}(x,x)
 \right)
\right\} \gamma_\mu \ ,
 \end{equation}

\begin{equation}
 \label{gapeq44}
(\Delta^+)^{44}  =  -i \frac{g^2 }{4 \Lambda^2} \gamma^\mu\left\{ -
\frac 13 \left( 2S^{88}_{(0)_{21}}(x,x) + \sqrt{2}
S^{89}_{(0)_{21}}(x,x) - 2 S^{99}_{(0)_{21}}(x,x)
 \right)
- \frac 23 S_{(+)_{21}}(x,x)
 \right\} \gamma_\mu  \ ,
\end{equation}

\begin{equation} \label{gapeq88}
(\Delta^+)^{88}  =  -i \frac{g^2}{4 \Lambda^2} \gamma^\mu\left\{
S^{11}_{(0)_{21}} (x,x) + \frac 13 \left(S^{88}_{(0)_{21}}(x,x) - 2
\sqrt{2} S^{89}_{(0)_{21}}(x,x) + 2 S^{99}_{(0)_{21}}(x,x)
 \right)
- \frac 83 S_{(+)_{21}}(x,x)
 \right\} \gamma_\mu  \ ,
\end{equation}

\begin{equation}
\label{gapeq99} (\Delta^+)^{99}  =  - i\frac{g^2}{4 \Lambda^2}
\gamma^\mu\left\{ 2 S^{11}_{(0)_{21}} (x,x) + \frac 23
S^{88}_{(0)_{21}}(x,x) + \frac 83 S_{(+)_{21}}(x,x)
 \right\} \gamma_\mu  \ ,
\end{equation}
\begin{equation}
 \label{gapeq89}
(\Delta^+)^{89}  =  -i \frac{g^2 \sqrt{2}}{12\Lambda^2}
\gamma^\mu\left\{3 S^{11}_{(0)_{21}} (x,x) -
\left(S^{88}_{(0)_{21}}(x,x) + \sqrt{2} S^{89}_{(0)_{21}}(x,x)
\right) -  2 S_{(+)_{21}}(x,x)
 \right\} \gamma_\mu  \ .
\end{equation}
\end{subequations}

From Eqs.~(\ref{gapCFLbasis}) one can immediately deduce

\begin{subequations}
\begin{eqnarray}
\Delta_A & = & - \frac 16 \left( 5 \Delta_{11} - \Delta_{88} - \Delta_{99} \right)  \ ,\\
\Delta_S & = &  \frac 16 \left(  \Delta_{11} + \Delta_{88} + \Delta_{99} \right) \ , \\
\Delta_N & \equiv & \Delta_S^B+\Delta_A^B = \frac{1}{12} \left(
\Delta_{11} -5 \Delta_{88} + 4
\Delta_{99} \right) \ , \\
\Delta_C & \equiv & \Delta_S^B-\Delta_A^B = \Delta_{44} \ .
\end{eqnarray}
\end{subequations}

Therefore,

\begin{subequations}
\begin{equation} \label{gapeqS}
\Delta^+_{S}  =  - i\frac{g^2 }{18\Lambda^2} \gamma^\mu\left\{
S^{11}_{(0)_{21}} (x,x) + S^{88}_{(0)_{21}} (x,x) +
S^{99}_{(0)_{21}} (x,x)
 \right\} \gamma_\mu  \ ,
\end{equation}

\begin{equation}
 \label{gapeqA}
\Delta^+_{A}  =  i \frac{g^2 }{12\Lambda^2} \gamma^\mu\left\{ -
\frac{17}{3} S^{11}_{(0)_{21}}  (x,x) + \frac 13 \left(
 S^{88}_{(0)_{21}} (x,x) + 6 \sqrt{2} S^{89}_{(0)_{21}}  (x,x) + 4 S^{99}_{(0)_{21}} (x,x) \right)
 \right\} \gamma_\mu  \ ,
\end{equation}

\begin{equation}
\label{gapeqSB} (\Delta^B_{S})^+  =  - i\frac{g^2 }{72\Lambda^2}
\gamma^\mu\left\{  S^{11}_{(0)_{21}}  (x,x) - 5 S^{88}_{(0)_{21}}
(x,x)   + 4 S^{99}_{(0)_{21}} (x,x)
 + 12  S_{(+)_{21}} (x,x)
 \right\} \gamma_\mu  \ ,
\end{equation}

\begin{equation}
\label{gapeqAB} (\Delta^B_{A})^+  =  - i\frac{g^2 }{72\Lambda^2}
\gamma^\mu\left\{ S^{11}_{(0)_{21}}  (x,x) + 7 S^{88}_{(0)_{21}}
(x,x) + 6 \sqrt{2} S^{89}_{(0)_{21}} (x,x)- 8 S^{99}_{(0)_{21}}
(x,x) +
 24  S_{(+)_{21}} (x,x)
 \right\} \gamma_\mu  \ ,
\end{equation}
\end{subequations}

Using the explicit expressions of the propagators given in
Sec.~\ref{NG-propagators} we can write the gap equations in momentum
space. In the limit $\Delta_S \ll \Delta_A$, and dropping the
contribution of antiparticles, Eqs. (\ref{gapeqS}), (\ref{gapeqA}),
(\ref{gapeqSB}), and (\ref{gapeqAB}) are respectively given by

\begin{subequations}
\begin{equation} \label{symme-gapeq}
 \Delta_S =  i\frac {g^2}{9 \Lambda^2}  \int \frac{d^4
q}{(2 \pi)^4} \left\{  \frac{\Delta_A}{  q^2_0 - (|{\bf q}| -\mu)^2-
\Delta_A^2 } - \frac{ C_a + M}{ q^2_0 - (|{\bf q}| -\mu)^2 -
\Delta_a^2 } -  \frac{ C_b - M }{ q^2_0 - (|{\bf q}| -\mu)^2 -
\Delta_b^2 } \right\} \ ,
 \end{equation}

\begin{equation} \label{antisymme-gapeq}
 \Delta_A  =  i\frac {g^2}{9 \Lambda^2}  \int \frac{d^4
q}{(2 \pi)^4} \left\{ \frac{17}{2} \frac{\Delta_A}{ q^2_0 - (|{\bf
q}| -\mu)^2- \Delta_A^2 } + \frac 12\frac{7 C_a - 2M}{ q^2_0 -
(|{\bf q}| -\mu)^2 - \Delta_a^2 } + \frac 12 \frac{7 C_b + 2M}{
q^2_0 - (|{\bf q}| -\mu)^2 - \Delta_b^2 } \right\} \ ,
 \end{equation}

\bea \label{BS-gapeq}
 \Delta^{B}_S &  = &-i\frac {g^2}{36 \Lambda^2} \left\{ \int \frac{d^4
q}{(2 \pi)^4} \left\{ \frac{-\Delta_A}{  q^2_0 - (|{\bf q}| -\mu)^2-
\Delta_A^2 } + \frac{F_a + 9 D_a}{ q^2_0 - (|{\bf q}| -\mu)^2 -
\Delta_a^2 } + \frac{F_b + 9 D_b}{ q^2_0 - (|{\bf q}| -\mu)^2 -
\Delta_b^2 } \right\}
\right. \nonumber \\
 & + & \left.
  12 \tilde {e} \tilde {B} \sum_{k=0}^{\infty} \int \frac{d q_0 d
q_3}{(2 \pi)^3} \frac{\Delta_C }{ q^2_0 - (|{\bf \bar q}| -\mu)^2 -
\Delta_C^2 } \right\} \ , \eea

\bea \label{BA-gapeq}
 \Delta^{B}_A & = &   -i\frac {g^2}{36 \Lambda^2} \left\{ \int \frac{d^4 q}{(2 \pi)^4} \{
\frac{-\Delta_A}{  q^2_0 - (|{\bf q}| -\mu)^2- \Delta_A^2 } +
\frac{F_a -9 D_a}{ q^2_0 - (|{\bf q}| -\mu)^2 - \Delta_a^2 } +
\frac{F_b -9 D_b}{ q^2_0 - (|{\bf q}| -\mu)^2 -
\Delta_b^2 } \right. \nonumber \\
 & + & \left.
  24 \tilde {e} \tilde {B} \sum_{k=0}^{\infty} \int \frac{d q_0 d
q_3}{(2 \pi)^3} \frac{\Delta_C }{ q^2_0 - (|{\bf \bar q}| -\mu)^2 -
\Delta_C^2 } \right\} \ , \eea
where $\Delta^2_{a/b}$ was defined in Eq.~(\ref{gaps-ab}), and
\begin{equation}\label{Cab}
 C_{a/b} = \frac{\Delta_A}{2} \left( 1
\pm \frac{4 \Delta^2_N + \Delta^2_A}{\Delta_A \sqrt{\Delta^2_A + 8
\Delta^2_N }} \right) \ , \qquad  M =  \frac{ 2 \Delta^2_N }{
\sqrt{\Delta^2_A + 8 \Delta^2_N }} \ ,
\end{equation}
 \begin{equation}\label{Fab} F_{a/b} = \frac{\Delta_A - 6 \Delta_N}{2}  \pm \frac{
\Delta_A^2 -4 \Delta_N^2 - 6 \Delta_N \Delta_A}{2 \sqrt{\Delta^2_A +
8 \Delta^2_N }} \ ,
 \end{equation}
%
\begin{equation}\label{Dab} D_{a/b} = \Delta_N \left( 1 \pm \frac{ \Delta_A}{
\sqrt{\Delta^2_A + 8 \Delta^2_N }}  \right) \ .
\end{equation}

\end{subequations}

These are highly coupled non-linear equations, that can only be
studied numerically. However, one can find a limiting situation
where an analytical solution can be found. It corresponds to the
situation of moderately high ${\widetilde B}$ fields $(\mu^2 <
\widetilde e \widetilde B<\Lambda^2)$.

\subsection{Zero magnetic field limit}

For completion we show here that in the limit of  zero magnetic
field the gap equations (\ref{symme-gapeq})-(\ref{BA-gapeq})
reduce, as it should be, to the CFL gap equations.

In the ${\tilde B} \rightarrow 0$ limit, one should have
$\Delta_{A/S}^B \rightarrow \Delta_{A/S}$. Using

\begin{equation}\label{CFL-limit}
\Delta_{A}^{B}=\Delta_{A} \gg \Delta_{S}^{B}=\Delta_{S},
\end{equation}
in the definitions (\ref{gaps-ab}) and (\ref{Cab})-(\ref{Dab}),
they take a more simplified form given by

\begin{eqnarray}\label{CFL-Param}
 \qquad \Delta^2_a  \approx 4
\Delta^2_A \ , \qquad \Delta^2_b \approx\Delta^2_A \ , \qquad
C_a   \approx  \frac 43 \Delta_A \ , \qquad  C_b \approx - \frac 13 \Delta_A \ ,\qquad\qquad\qquad\, \nonumber \\
F_a \approx -4 \Delta_A, \qquad F_b \approx -\Delta_A, \qquad D_a
\approx \frac {4} {3} \Delta_A, \qquad D_b \approx \frac {2} {3}
\Delta_A, \qquad M \approx \frac 23 \Delta_A \ .\qquad
\end{eqnarray}

Making use of (\ref{CFL-limit}) and (\ref{CFL-Param}) in the right
hand side of Eqs.~(\ref{symme-gapeq}-\ref{BA-gapeq}) and taking
into account that when ${\tilde B} \rightarrow 0$ the Landau level
summation should be replaced by an integral following the
prescription

\begin{equation}\label{LandauSum} \tilde e \tilde B \sum_{n=0}^{\infty} f(2\tilde e \tilde B
n)\rightarrow \int_{-\infty}^{\infty} \frac{dq_{1}dq_{2}}{2 \pi}
f(q_{\bot}^2)
\end{equation}
we can easily see that both Eqs.~(\ref{symme-gapeq}) and
(\ref{BS-gapeq}) reduce to the CFL gap equation for the symmetric
gap

\begin{subequations}
\begin{equation}
\label{CFLsymme-gapeq}
 \Delta_S  =   i \frac {g^2}{9 \Lambda^2}  \int \frac{d^4
q}{(2 \pi)^4} \left\{  \frac{2 \Delta_A}{ q^2_0 - (|{\bf q}|
-\mu)^2 - \Delta_A^2 } - \frac{2 \Delta_A }{ q^2_0 - (|{\bf q}|
-\mu)^2 -4\Delta_A^2 } \right\} \ ,
\end{equation}
while both Eqs. (\ref{antisymme-gapeq}) and (\ref{BA-gapeq})
reduce to the CFL gap equation for the antisymmetric gap

\begin{equation}
\label{CFLantisymme-gapeq} \Delta_A =  i \frac {g^2}{9 \Lambda^2}
\int \frac{d^4 q}{(2 \pi)^4} \left\{ \frac{ 8 \Delta_A}{ q^2_0 -
(|{\bf q}| -\mu)^2 - \Delta_A^2 } +  \frac{4 \Delta_A }{ q^2_0 -
(|{\bf q}| -\mu)^2 -4\Delta_A^2 } \right\} \ .
\end{equation}
\end{subequations}

Therefore, starting from the
Eqs.~(\ref{symme-gapeq})-(\ref{BA-gapeq}) we recover the correct CFL
gap equations in the NJL theory. The quark propagators in the CFL
case can be straightforwardly found, (see for example Ref.
\cite{Litim:2001mv}), assuming that $\Delta_S \ll \Delta_A$. The
solution reads ($\delta \equiv \Lambda - \mu$)
\begin{equation}
\label{CFL-Agap} \Delta_A^{\rm CFL} \sim 2 \sqrt{\delta \mu} \,\exp{
\Big(-\frac{ 3 \Lambda^2 \pi^2}{ 2 g^2 \mu^2}\Big)} \ ,
\end{equation}
and
\begin{equation} \Delta_S^{\rm CFL} \sim \Delta_A^{CFL}
\frac{g^2 \mu^2}{9 \Lambda^2 \pi^2} \ln{2} \ .
\end{equation}

\subsection{Strong magnetic field limit}

Let us now consider the MCFL gap equations for the case
$\widetilde{e}\widetilde{B} \gtrsim \mu^2$. Taking into account that
the main contribution to the gap comes from quark energies near the
Fermi level, it follows that in the strong field region the leading
contribution comes from the lowest Landau level (LLL).

Assuming $\Delta^B_A \gg \Delta^B_S, \Delta_A$, and $\Delta_A \gg
\Delta_S$, the definitions (\ref{gaps-ab}),
(\ref{Cab})-(\ref{Dab}) reduce to

\begin{eqnarray}\label{MCFL-Param}
\qquad \Delta^2_a  \approx 2(\Delta^B_A)^2 \ , \qquad \Delta^2_b
\approx 2(\Delta^B_A)^2 \ , \qquad
C_{a/b}  \approx  \frac{\Delta_{A}}{2}\pm \frac{\sqrt{2}}{2} \Delta^B_A \ , \qquad \qquad\qquad\qquad\, \nonumber \\
F_a \approx -(3+\frac{\sqrt{2}}{2}) \Delta^B_A, \qquad F_b \approx
-(3-\frac{\sqrt{2}}{2}) \Delta^B_A, \qquad D_a \approx \Delta^B_A,
\qquad D_b \approx \Delta^B_A, \qquad M \approx \frac{\sqrt{2}}{2}
\Delta^B_A \ .\qquad
\end{eqnarray}

In this approximation the gap equations decouple and the equation
(\ref{BA-gapeq}) for $\Delta^B_A$ becomes

\begin{equation} \label{maingeq} \Delta^B_A  \approx  \frac{g^2}{3 \Lambda^2}
\int_{\Lambda} \frac{d^3 q}{(2 \pi)^3} \frac{
\Delta^B_A}{\sqrt{(|{\bf q}| -\mu)^2 + 2 (\Delta^B_A)^2 }}
 +  \frac{g^2 \widetilde{e}\widetilde{B}}{3 \Lambda^2} \int_{-\Lambda}^{\Lambda}
\frac{d q_{3}}{(2 \pi)^2} \frac{ \Delta^B_A}{\sqrt{(|q_{3}|-\mu)^2 +
(\Delta^B_A)^2 }}   \ . \end{equation}

Its solution reads
\begin{equation}
\label{gapBA} \Delta^B_A \sim 2 \sqrt{\delta \mu} \, \exp{\Big( -
\frac{3 \Lambda^2 \pi^2} {g^2 \left(\mu^2 + \widetilde{e}
\widetilde{B} \right)} \Big) } \ ,
\end{equation}

The exponent in (\ref{gapBA}) has the typical BCS form $\exp
~[~1/N\widetilde{G}~]$, where $\widetilde{G}=g^2/3\Lambda^{2}$ is
the characteristic effective coupling constant of the
$\overline{3}$ channel \cite{Son}, and $N$ represents the total
density of states at the Fermi surface of the four quasiquarks of
a single chirality contributing to the antisymmetric gap parameter
$\Delta_{A}^{B}$. If no magnetic field is present (CFL case),
$N=4N_{\mu}$, where
 $N_{\mu}=\mu^2/2\pi^2$ is the density of states of
one quasiquark. At nonzero magnetic field the density splits in two
terms $N=2N_{\mu} +2N_{\widetilde{B}}$, because two of the four
quasiquarks are neutral, thus they have density $N_{\mu}$, while two
are charged and their density of states depends on the magnetic
field. In the zero Landau level the density of states of a charged
quasiquark is $N_{\widetilde{B}}=\widetilde{e}\widetilde{B}/2\pi^2$.
Notice that the gap parameter (\ref{gapBA}) increases with the
magnetic field. Comparing Eqs.(\ref{CFL-Agap}) and (\ref{gapBA}) one
can see that $\Delta_{A}^{B}>\Delta_{A}^{CFL}$ if
$\widetilde{e}\widetilde{B}> \mu^{2}$, as assumed in the derivation
of Eq.(\ref{maingeq}). This means that the field not just changes
the gap structure, but it also enhances the gaps that get
contributions from pairs formed by charged quarks.

As mentioned in the Introduction, although this situation has some
similarity with the magnetic catalysis of chiral symmetry breaking
\cite{MC}; the way the field influences the pairing mechanism in
the two cases is quite different. The particles participating in
the chiral condensate are near the surface of the Dirac sea. The
effect of a magnetic field there is to effectively reduce the
dimension of the particles at the lowest Landau level, which in
turn strengthens their effective coupling, catalyzing the chiral
condensate. Color superconductivity, on the other hand, involves
quarks near the Fermi surface, with a pairing dynamics that is
already $(1+1)$-dimensional. Therefore, the ${\widetilde B}$ field
does not yield further dimensional reduction of the pairing
dynamics near the Fermi surface and hence the lowest Landau level
does not have a special significance here. Nevertheless, the field
modifies the density of states of the ${\widetilde Q}$-charged
quarks, and it is through this effect, as shown in Eq.
(\ref{gapBA}), that the pairing of the charged particles is
reinforced by a penetrating strong magnetic field.

In the strong field approximation the remaining gap equations
(\ref{symme-gapeq})-(\ref{BS-gapeq}) are respectively given by
\begin{equation} \label{symme-gapeqapp}
 \Delta_S \approx   \frac {g^2}{18 \Lambda^2}  \int_{\Lambda} \frac{d^3q}{(2 \pi)^3}
\Big(  \frac{\Delta_A}{ \sqrt{ (|{\bf q}| -\mu)^2 + \Delta_A^2 }}
 -    \frac{\Delta_A}{ \sqrt{(|{\bf q}| -\mu)^2 + 2 (\Delta^B_A)^2 } }
\Big) \ ,
 \end{equation}
\begin{equation} \label{antisymme-gapeqapp}
 \Delta_A  \approx  \frac {g^2}{4 \Lambda^2}  \int_{\Lambda} \frac{d^3q}{(2 \pi)^3}
 \Big( \frac {17}{9} \frac{\Delta_A}{ \sqrt{ (|{\bf q}| -\mu)^2 + \Delta_A^2 }}
 +   \frac{7}{9} \frac{\Delta_A}{ \sqrt{(|{\bf q}| -\mu)^2 + 2 (\Delta^B_A)^2 }  }
 \Big) \ ,
 \end{equation}
and
\begin{equation}
\label{symgeq} \Delta^B_S  \approx  -\frac{g^2}{6 \Lambda^2}
\int_{\Lambda} \frac{d^3 q}{(2 \pi)^3} \frac{
\Delta^B_A}{\sqrt{(|{\bf q}| -\mu)^2 + 2 (\Delta^B_A)^2 }}
 +  \frac{g^2 \widetilde{e}\widetilde{B}}{6 \Lambda^2} \int_{-\Lambda}^{\Lambda}
\frac{d q_{3}}{(2 \pi)^2} \frac{ \Delta^B_A}{\sqrt{(|q_{3}|-\mu)^2 +
(\Delta^B_A)^2 }}   \ , \end{equation}

The corresponding solutions are

\begin{equation}\label{symm-gapsol}
\Delta_S \simeq \frac{2}{17} (1-\frac{2}{1+y})\Delta_{A},
\end{equation}

\begin{equation}\label{antisymm-gapsol}
\Delta_{A} \simeq \sqrt{4\mu\delta}
\exp[-\frac{36}{17x}+\frac{21}{17x(1+y)}],
\end{equation}

\begin{equation}\label{symm-Bgapsol}
\Delta_{S}^{B} \simeq (\frac{y}{1+y}-\frac{1}{2})\Delta_{A}^{B},
\end{equation}
where $x \equiv g^2 \mu^2/\Lambda^2 \pi^2$, and $y \equiv
\widetilde{e}\widetilde{B}/\mu^2$. As expected, the solutions
(\ref{gapBA}),(\ref{symm-gapsol})-(\ref{symm-Bgapsol}) are
consistent with the initial assumptions $\Delta^B_A \gg \Delta^B_S,
\Delta_A$, and $\Delta_A \gg \Delta_S$ if the field satisfies
$\widetilde{e}\widetilde{B}>\mu^{2}$. When the field increases
within this strong field region $\Delta_{A}^B$ grows and
$\Delta_{A}$ decreases, and they clearly split. Even for fields just
slightly larger than $\mu^{2}$ the results are reliable. For
example, for $y= 3/2$ and $x \sim 0.3$ \cite{Casalbuoni:2003cs}, one
finds $\Delta_A \sim 0.2\, \Delta^B_A$ for $y= 3/2$, while for $x
\sim 1$ then $\Delta_A \sim 0.5 \,\Delta^B_A$. For $\mu \sim 350-
400$ MeV, one can estimate the field strength for $y= 3/2$ to be
$\widetilde{e}\widetilde{B} \sim (1.2 - 1.6) \cdot 10^{18}$G.

\section{Effective field theory for the Goldstone bosons of the MCFL phase}

In the absence of a magnetic field, three-flavor massless quark
matter at high baryonic density is in the CFL phase on which
diquark condensates lock the $SU(3)$ color and $SU(3)$ flavor
transformations thereby breaking both symmetries according to the
following symmetry breaking pattern
\begin{equation}
 SU(3)_C \times SU(3)_L \times SU(3)_R \times U(1)_B
\rightarrow SU(3)_{C+L+R} \ .
\end{equation}
Nine Goldstone bosons appear as the result of the Anderson-Higgs
mechanism. One is a singlet, scalar mode, associated to the
breaking of the baryonic symmetry, and the remaining octet is
associated to the axial $SU(3)_A$ group, just like the octet of
mesons in vacuum.

Strictly speaking, once electromagnetic effects are taken into
account, the flavor symmetries $SU(3)_L \times SU(3)_R $ of the
original three-flavor massless quark matter are reduced to $SU(2)_L
\times SU(2)_R \times U(1)^{(1)}_A$ with $U(1)^{(1)}_A \subset
SU(3)_A$, since the electromagnetic charge of $d$ and $s$ quarks is
different from that of the $u$ quark. However, because the
electromagnetic structure constant $\alpha_{\rm e.m.}$ is so small,
this effect is really tiny, a small perturbation, and one can still
consider the original flavor as an approximated symmetry of the
theory.

On the other hand, we know that in the CFL phase a so-called
``rotated" electromagnetism remains, allowing the penetration of
the color superconductor by a conventional magnetic field in the
form of a ``rotated" magnetic field. If the penetrating magnetic
field is strong enough, the effect of electromagnetism cannot be
treated anymore as a small perturbation since, as shown in
previous sections, it affects the pairing phenomena and leads to
the development of the MCFL phase, whose symmetry breaking pattern
is

\begin{equation}
SU(3)_C \times SU(2)_L \times SU(2)_R \times U(1)^{(1)}_A\times
U(1)_B \times U(1)_{\rm e.m.} \rightarrow SU(2)_{C+L+R} \times
{\widetilde U(1)}_{\rm e.m.} \ .
\end{equation}
As mentioned in Section \ref{selfen-gen-struc} the locked $SU(2)$
group corresponds to the maximal unbroken symmetry, such that it
maximizes the condensation energy. The counting of broken generators
here, after taking into account the Anderson-Higgs mechanism, tells
us that there are only five Goldstone bosons, all of which are
neutral with respect to the rotated charge. As in the CFL case, one
is associated to the breaking of the baryon symmetry; three
Goldstone bosons are associated to the breaking of $SU(2)_A$, and
another one is associated to the breaking of $U(1)^{(1)}_A$.
Therefore, apart from modifying the structure and magnitude of the
gap, the applied strong  magnetic field does also affect the low
energy properties of the color superconductor, since it reduces the
number of Goldstone bosons from nine to five, none of which is
charged.

Following an approach similar to the one discussed in
\cite{son-stephanov-prd61-074012} and \cite{Casalbuoni:1999wu} we
can derive the low-energy effective theory for the Goldstone bosons
in the MCFL phase. Here we will just sketch the main steps leaving
detailed calculations for a future work. The main idea is to
introduce left and right coset fields
\begin{equation}
X^{ia} \sim \epsilon^{ijk} \epsilon^{abc} \langle \psi^{bj}_L
\psi^{ck}_L \rangle^* \ , \qquad Y^{ia} \sim \epsilon^{ijk}
\epsilon^{abc} \langle \psi^{bj}_R \psi^{ck}_R \rangle^*
\end{equation}
where $a,b,c$ denote flavor indices, $i,j,k$ denote color indices,
and $L/R$ denote left/right chirality, respectively; which transform
under the global transformations of the original group as
\begin{equation}
X \rightarrow U_L X U_C^\dagger \ , \qquad Y \rightarrow U_R Y
U_C^\dagger \ .
\end{equation}
where $U_{C}$ is a color transformation, while $U_L \in
SU(2)_{L}\times U(1)^{(1)}_L$ and $U_R \in SU(2)_{R} \times
U(1)^{(1)}_R$ are left and right flavor transformations
respectively. We are ignoring in our discussion the baryon symmetry,
but its Goldstone mode can be incorporated in a similar way.

Then, keeping in mind that the Goldstone modes are related to
motions along the degenerate minima of the effective potential, one
can obtain the effective action for the Goldstone modes just by
factoring out the modulus of the condensate and allowing the phases
of $X$ and $Y$ to depend on time and space.

At this point it is convenient to introduce the flavor singlet
\begin{equation}
\Sigma = X Y^\dagger= \exp{ \left(i \frac{\Phi}{f_{\pi, B}} + i
\phi_0 \right) } \ ,  \qquad \Phi = \phi_A \sigma^A  \ , \qquad A=
1,2,3.
\end{equation}
where $\sigma_A$ are the $SU(2)$ generators and the fields $\phi_A$
are associated to the breaking of $SU(2)_A$, while $\phi_0$ is
associated to the breaking of $U(1)^{(1)}_A$.

Taking into account that the effective low energy Lagrangian of the
Goldstone bosons in the MCFL should be symmetrical under the
rotations of the original group $SU(3)_{C} \times SU(2)_L \times
SU(2)_R \times U(1)^{(1)}_A $, we can write it down up to leading
order in the derivatives as
\begin{equation}
\label{chiral-lag-MCFL} {\cal L}   = \frac{f_{\pi,B}^2}{4} \left(
{\rm Tr} \left( \partial_0 \Sigma
\partial_0 \Sigma^\dagger \right) +
\left( v^2_\perp g^{i j}_\perp + v^2_\parallel g^{i j}_\parallel
\right) {\rm Tr} \left( \partial_i \Sigma
\partial_j \Sigma^\dagger \right) \right) \ .
\end{equation}

Notice that we introduced different longitudinal and transverse
velocities. This should be done because besides the usual Lorentz
symmetry breaking proper of the theory at finite density, the strong
magnetic field induces and extra symmetry reduction, since only the
spacial $SO(2)$ rotations in the plane perpendicular to the magnetic
field are allowed. The constants $f_{\pi,B}$, $v_\perp$ and
$v_\parallel$ should be computed from the microscopic theory. We
leave such a computation for a future project.

It is remarkable that although the external magnetic field does not
explicitly enter in the low-energy Lagrangian
(\ref{chiral-lag-MCFL}), since it does not couple minimally to the
MCFL Goldstone bosons because these are all neutral with respect to
the rotated charge, its presence is manifested anyway through the
anisotropy of the velocities.

The effective field theory for these Goldstone fields is totally
analogous to that found for QCD in a strong magnetic field in Ref.
\cite{miransky-shovkovy-02}. Were it not for the presence of the
extra Goldstone boson associated to the baryon symmetry breaking in
the MCFL phase, the two low-energy theories, at low and high
densities, in an external magnetic field would be completely
equivalent, and one would expect them to be connected by a
crossover. We might think that because the existence of the extra
Goldstone mode, this is not possible. However, the situation is
similar to the case without magnetic field. There also the high and
low density effective theories of the Goldstone modes are identical
with the exception of the extra Goldstone mode in the color
superconductor due to baryon symmetry breaking, and yet it has been
argued \cite{schafer-Wilczek-prl82} that the extra mode in the CFL
phase may have a counterpart in the chiral phase, the dibaryon state
$H$ found in Ref. \cite{jaffe-prl38}. Therefore, the question about
the possible equivalence between high and low density low energy
theories in a magnetic field still remains open.

\section{Concluding Remarks}

In this paper we investigated the effect of a magnetic field on the
color superconducting gap of a dense, three massless quark flavor
system. Our main result is that the magnetic field leads to the
formation of a new color-flavor locking phase, that we have called
the MCFL phase. We found that the magnetic field affects essentially
the gap by modifying the density of states of the charged quarks on
the Fermi surface.

In the analytic calculations performed in this paper we assumed a
strong field approximation ($\widetilde{e}\widetilde{B} \gtrsim
\mu^2$), which corresponds to fields $\sim 10^{18}$G. The only
reason we considered such strong fields was to simplify our task so
we could avoid to sum in Landau levels, and hence could perform
analytical calculations. However, we underline that there is nothing
special about the LLL here, because the main effect of the field is
to modify the density of states and that is not restricted only to
the LLL. On the other hand, there are indications, based on the
study of the low energy degrees of freedom of the CFL effective
field theory in the presence of an external magnetic field
\cite{CM-QFTExt}, that for ${\tilde e} {\tilde B}^{\rm ext} \sim 5
\cdot 10^{16}$ G the symmetry pattern of the theory qualitatively
separates from that of the CFL phase and becomes characteristic of
the new MCFL phase. These estimates suggest that a distinguishable
separation between the gaps $\Delta_A$ and $\Delta^B_A$ may take
place already at those magnetic field orders. Of course, a
definitive determination of the field strength required to separate
the two phases can only be obtained through a numerical calculation
on which the effect of all the higher Landau levels in the gap
equations are included.

We have shown that the MCFL phase has a smaller vector symmetry than
the CFL phase and consequently the number of Goldstone bosons
reduces from 9 to 5. Our zero-temperature results imply that a
propagating rotated photon with energy less than the lightest
charged quark mode cannot scatter, since all the
$\widetilde{Q}$-charged quarks acquire a gap and all the
Nambu-Goldstone bosons are neutral. The anisotropy present in the
background of an external magnetic field and the existence of
charged Goldstone bosons in CFL but not in MCFL indicates a rather
different low energy physics in these two phases and it should be
reflected in the transport properties. In particular, the MCFL
superconductor is transparent and behaves at $T=0$ as an anisotropic
dielectric, as opposed to the isotropic dielectric behavior of the
CFL phase \cite{Litim:2001mv, Manuel:2001mx}. However, similar to
the CFL, the medium will become optically opaque as soon as leptons
are thermally excited \cite{Shov-Ellis03}.

Although the $\widetilde{Q}$ neutrality of the MCFL is guaranteed
without having to introduce any electron density due to the absence
of gapless modes and the $\widetilde{Q}$-neutrality of the diquark
condensates, imposing the 3- and 8- color neutralities may require
the introduction of small "chemical potentials" $\mu_{3}$ and
$\mu_{8}$ due to the difference between the main gaps $\Delta_{A}$
and $\Delta_{A}^{B}$.

If we introduce quark masses the scenario will be much more
complicated though. First, we would have to revise the neutrality
conditions and see how they are affected. In addition, one would
expect that the effect of the magnetic field increasing the gaps
formed by $\widetilde{Q}$-charged quarks might have some impact in
counteracting the stress produced by the mass of the strange quark
on the corresponding Fermi sphere and perhaps also in the onset of
gapless modes, along with the problem of the instability of the
gluon modes due to imaginary Meissner masses. Considering the
consequences of incorporating quark masses, together with a careful
study of the effects of the magnetic field in the low energy
physics, in transport properties or in neutrino dynamics will be the
subject of future investigations.

Even if only at a speculative level, it is worth to mention some
possible astrophysical consequences of the main results of this
paper. The equation of state of the color superconductor will be
affected by the magnetic field, as different quark gaps will be
affected in a different way by the field. We do not expect this to
be a pronounced effect, though, but it might be interesting to see
whether it affects the mass-to-radius ratio of the star. Because the
low energy physics of the CFL and MCFL are so different, depending
on the strength of the star's magnetic field there may be different
signatures in the cooling process of the star. Transport properties,
such as viscosities or thermal conductivities, will be also affected
by the presence of the magnetic field. Finally, the dynamics of the
magnetic field itself might be very peculiar in quark matter,
differing from that in a neutron star, which is commonly believed to
be an electromagnetic superconductor.

\textbf{Acknowledgments}

This work  was supported in part by NSF grant PHY-0070986,  by MEC
under grants FPA2004-00996 and AYA 2005-08013-C03-02, and by GVA
under grant GV05/164.

\appendix

\section{Projectors in the presence of a background magnetic field}

\label{App-A}

In the absence of a background magnetic field  the condensate  is usually given in terms
of different projectors \cite{Pisarski-rischke83-1999}. The chiral and helicity projection operators are given respectively by
\begin{equation}
P_{R,L}=\frac{1\pm \gamma _{5}}{2} ,
\label{chiral-proj}
\end{equation}
and
\begin{equation}
H_{\pm }(p)=\frac{1}{2}[1\pm \gamma _{5}\gamma
_{0}\overrightarrow{\gamma }\cdot \widehat{\overrightarrow{p}}] \ .
\label{helicityOperator}
\end{equation}
The massless quark free-energy projector for positive
and negative energies is given by
\begin{equation}
\Lambda ^{\pm }(p)=\frac{1\pm \gamma _{0}\overrightarrow{\gamma
}\cdot \widehat{\overrightarrow{p}}}{2},
 \label{free-energy-proj}
\end{equation}

For massless quarks, the chiral and energy projectors are enough to
specify the quark propagators. Furthermore, in a NJL theory, where
there are only contact interactions, the gap is a momentum independent constant, and
the Dirac structure of the condensate¡ is particularly simple in the spin zero case,
where it simply reduces to $C\gamma_5$, where $C = i \gamma_2 \gamma_0$ is the matrix of charge conjugation.

In the presence of an external $\widetilde{B}$ field
the helicity-projection operators (\ref{helicityOperator}) and the
free-energy projectors (\ref{free-energy-proj}) are not conserved quantities,
since they do not commute with the field-dependent Hamiltonian. But
their generalization in terms of the field dependent momentum
operator (\ref{pi-operator}) are.  In covariant form, these conserved
operators can be expressed respectively by

\begin{equation}
\widetilde{H}_{\pm }(\Pi )=\frac{1\pm \frac{i}{2}\gamma _{5}[u_{\mu }%
\widehat{\overrightarrow{\Pi }}_{\nu }-u_{\nu
}\widehat{\overrightarrow{\Pi }}_{\mu }]\sigma ^{\mu \nu }}{2} \ ,
\label{18}
\end{equation}

\begin{equation}  \label{26}
\widetilde{\Lambda }^{\pm }(\Pi)=\frac{1\pm \frac{i}{2}[u_{\mu }%
\widehat{\overrightarrow{\Pi }}_{\nu }-u_{\nu
}\widehat{\overrightarrow{\Pi }}_{\mu }]\sigma ^{\mu \nu }}{2}    \ ,
\end{equation}
where $u^\mu$ is the four-velocity of the center of mass of the many particle system.
In the rest frame, the helicity and energy projectors can be
expressed in momentum space respectively as

\begin{equation}  \label{27}
\widetilde{H}^{(\pm)}_{\pm }(\overline{p})=\frac{1\pm \gamma
_{5}\gamma_{0}\gamma\cdot
\widehat{\overline{\mathbf{p}}}^{(\pm)}}{2} \ ,
\end{equation}

\begin{equation}  \label{rest-f}
\widetilde{\Lambda }_{(\pm)}^{\pm }(\overline{p})=\frac{1\pm\gamma_{0}\gamma\cdot
\widehat{\overline{\mathbf{p}}}^{(\pm)}}{2} \ .
\end{equation}

Similarly to the free case, in the presence of a ``rotated''
magnetic field the chiral (\ref{chiral-proj}), helicity
(\ref{helicityOperator}) and energy (\ref{rest-f}) projectors
commute.


\begin{thebibliography}{}

\bibitem{MCFL}
   E.~J.~Ferrer, V.~de la Incera and C.~Manuel,
  Phys.\ Rev.\ Lett.\  {\bf 95}, 152002 (2005).

\bibitem{reviews}
  K.~Rajagopal and F.~Wilczek,
 hep-ph/0011333; M.~Alford, Ann.\ Rev.\ Nucl.\ Part.\ Sci.\ {\bf 51}, 131 (2001);
G.~Nardulli,  Riv.\ Nuovo Cim.\  {\bf 25N3}, 1 (2002); T. Sch\"afer,
"Quark Matter," hep-ph/0304281; D.~H.~Rischke, Prog.\ Part.\ Nucl.\
Phys.\ {\bf 52}, 197 (2004); H.-C.~Ren, hep-ph/0404074;
I.~A.~Shovkovy, Found.Phys. {\bf 35} (2005) 1309-1358.

\bibitem{Witten}
 N. Itoh, Prog. Theor. Phys. {\bf 44}, 291 (1970);
A.~R.~Bodmer,
Phys.\ Rev.\ D {\bf 4}, 1601 (1971);
 E.~Witten,
Phys.\ Rev.\ D {\bf 30}, 272 (1984).

\bibitem{qstars}
M. Dey, I. Bombaci, J. Dey, S. Ray and C. B. Samanta, Phys. Lett.
{\bf B 438}, 123 (1998); X. D. Li, I. Bombaci, M. Dey, J. Dey, and
E. P. J. van den Heuvel, Phys. Rev. Lett. {\bf B 83}, 3776 (1999);
X. D. Li, S. Ray, J. Dey, M. Dey, and I. Bombaci, Ap. J. {\bf 527},
L51 (1999); F.~Weber,
  Prog.\ Part.\ Nucl.\ Phys.\  {\bf 54}, 193 (2005);
 Sinha M., M. Dey, S. Ray, and J. Dey; "Strange Stars and
Superbursts at near Eddington Mass Accreation Rates,"
astro-ph/0504292; D. Page, and A. Cumming, "Superbursts from strange
stars," astro-ph/0508444.


\bibitem{Grasso}
I. Fushiki, E. H. Gudmundsson, and C.J. Pethick, Astrophys. J. 342,
958 (1989); T.A. Mihara, et. al., Nature (London) 346, 250 (1990);
G. Chanmugam, Ann. Rev. Astron. Astrophys. 30, 143 (1992); P. P.
Kronberg, Rep. Prog. Phys.\textbf{57} 325 (1994); D.~Lai, Rev.\
Mod.\ Phys. {\bf 73}, 629 (2001); D. Grasso and H.R. Rubinstein,
Phys. Rep.\textbf{348} 163 (2001).

\bibitem{magnetars}
  C.~Thompson and R.~C.~Duncan,
  Astrophys.\ J.\  {\bf 473}, 322 (1996).

\bibitem{virial}
L. Dong and S.L. Shapiro, ApJ.\ {\bf 383}, 745 (1991).


\bibitem{alf-raj-wil-99/537}
M. Alford, K. Rajagopal and F.
Wilczek, Nucl. Phys. B \textbf{537}, 443 (1999).

\bibitem{alf-berg-raj-NPB-02}
M. Alford, J. Berges, and K. Rajagopal, Nucl. Phys. B \textbf{571},
269 (2000).

\bibitem{MC}
V.~P.~Gusynin, V.~A.~Miransky, and I.~A.~Shovkovy, \prl {\bf 73},
3499 (1994); \pl B {\bf 349}, 477 (1995); \prd {\bf 52}, 4747
(1995); Nucl. Phys. B {\bf 462}, 249 (1996); K.~G.~Klimenko, Z.
Phys. C{\bf 54}, 323 (1992); Teor. Mat. Fiz. {\bf 90}, 3 (1992).

\bibitem{orthonormality}
D.-S Lee, C. N. Leung and Y. J. Ng, \prd {\bf 55}, 6504 (1997); E.
J. Ferrer, and V. de la Incera, \prd {\bf 58}, 065008 (1998); \pl B
{\bf 481}, 287 (2000); E. Elizalde, E. J. Ferrer, and V. de la
Incera, \prd {\bf 68}, 096004 (2003).

\bibitem{Ritus:1978cj}
V.I. Ritus, Ann.Phys. 69, 555 (1972); Sov.\ Phys.\ JETP {\bf 48},
788 (1978) [Zh. Eksp. Teor. Fiz.{\bf 75}, 1560 (1978)].

\bibitem{alf-raj-JHEP-2002}  M. Alford, and K. Rajagopal,
JHEP \textbf {0206}, 031 (2002).

\bibitem{iida}
N.K. Glendenning, astro-ph/9706236; T.C. Phukon, Phys. Rev. D {\bf
62}, 023002 (2000);  E. V. Gorbar, Phys. Rev. D \textbf{62}, 014007
(2000); T. Ghosh, and S. Chakrabarty, Phys. Rev. D {\bf 63}, 043006
(2001); D. Ebert, K. G. Klimenko, H. Toki, and V. Ch. Zhukovsky,
Prog. Theor. Phys. {\bf 106}, 835 (2001); D.M. Sedrakian, D.
Blaschke, K.M. Shahabasyan and D.N. Voskresnsky, Astrofiz. {\bf 44}
(2001) 443; D.M. Sedrakian and D. Blaschke, Astrofiz. {\bf 45}
(2002) 203; D K. Iida and G. Baym, Phys. Rev. D \textbf{66}, 014015
(2002); D. Ebert, V. V. Khudyakov, V. Ch. Zhukovsky, and K. G.
Klimenko, Phys. Rev. D {\bf 65}, 054024 (2002); I. Giannakis and H-C
Ren, Nucl. Phys. B \textbf{669}, 462 (2003).

\bibitem{rajagop-schmitt-ph/0512043}K. Rajagopal and A. Schmitt,
Phys. Rev. D {\bf 73} (2006) 045003.

\bibitem{Pisarski-rischke83-1999}
R.~D.~Pisarski, and D.~H.~Rischke, \prl {\bf 83},
37 (1999); 
  Phys.\ Rev.\ D {\bf 60}, 094013 (1999);
  Phys.\ Rev.\ D {\bf 61}, 074017 (2000).

\bibitem{miransky-shovkovy-02} V.~A.~Miransky, and I.~A.~Shovkovy,
\prd {\bf 66}, 045006 (2002).

\bibitem{Bailin-Love}D.~Bailin, and A.~Love,
Phys. Rep. {\bf 107}, 325 (1984).

\bibitem{efi-ext}
E. Elizalde, E. J. Ferrer, and V. de la Incera, Ann. of Phys., {\bf
295}, 33 (2002); \prd {\bf 70}, 043012 (2004).

\bibitem{ayala-et-al-ph/0606209}A. Ayala, A. Bashir, A. Raya,
and E. Rojas, \textit{"Dynamical mass generation in strong coupling
QED with weak magnetic fields"}, hep-ph/0602209.

\bibitem{son-stephanov-prd61-074012} D.T. Son and M.A. Stephanov,Phys. Rev. D {\bf
61}, 074012 (2000).

\bibitem{Litim:2001mv}
D.~F.~Litim and C.~Manuel, Phys.\ Rev.\ D {\bf 64}, 094013 (2001).

\bibitem{Son}
D. T. Son, \prd {\bf 59}, 094019 (1999).

\bibitem{Casalbuoni:2003cs}
  R.~Casalbuoni, R.~Gatto, G.~Nardulli and M.~Ruggieri,
  \prd  {\bf 68}, 034024 (2003).



\bibitem{Casalbuoni:1999wu} R.~Casalbuoni and R.~Gatto,
Phys.\ Lett.\ {\bf B464} 111  (1999). 

\bibitem{schafer-Wilczek-prl82} T. Sch\"{a}fer and F.
Wilczek, \prl {\bf 82}, 3956 (1999).

\bibitem{jaffe-prl38} R. Jaffe, \prl {\bf 38}, 195 (1977), 617(E)
(1977).

\bibitem{CM-QFTExt} C.~Manuel, ``Low energy properties of color-flavor locked
superconductors," hep-ph/0512054;
C.~Manuel and M.~H.~Tytgat,
Phys.\ Lett.\ B {\bf 501}, 200 (2001). 


\bibitem{Manuel:2001mx}C.~Manuel and K.~Rajagopal, Phys.\ Rev.\ Lett.\  {\bf 88}, 042003 (2002).

\bibitem{Shov-Ellis03}I.~A.~Shovkovy and P.J. Ellis,
\prc {\bf 67}, 048801 (2003).

\end{thebibliography}
\end{document}